\documentclass[aps,pra,floatfix,footinbib,longbibliography,twocolumn,superscriptaddress, preprintnumbers, nobibnotes]{revtex4-1}

\usepackage{amsmath}
\usepackage{amsfonts}
\usepackage{amssymb}
\usepackage{graphicx}
\usepackage{color}
\usepackage{accents}
\usepackage[colorlinks=true, citecolor=blue]{hyperref}
\usepackage[all]{hypcap}
\usepackage{subfigure}
\usepackage[normalem]{ulem}
\newcommand{\ket}[1]{\left|#1\right\rangle}
\newcommand{\bra}[1]{\left\langle #1\right|}

\newcommand{\bea}{\begin{eqnarray}}
\newcommand{\eea}{\end{eqnarray}}
\newcommand{\be}{\begin{equation}}
\newcommand{\ee}{\end{equation}}
\newcommand{\ba}{\begin{align}}
\newcommand{\ea}{\end{align}}

\newcommand{\phd}{{\phantom\dagger}}

\makeatletter % the lines below make the double bra and double ket symbols
\newsavebox{\@brx}
\newcommand{\llangle}[1][]{\savebox{\@brx}{\(\m@th{#1\langle}\)}%
  \mathopen{\copy\@brx\kern-0.5\wd\@brx\usebox{\@brx}}}
\newcommand{\rrangle}[1][]{\savebox{\@brx}{\(\m@th{#1\rangle}\)}%
  \mathclose{\copy\@brx\kern-0.5\wd\@brx\usebox{\@brx}}}
\makeatother

\newlength{\dhatheight} % lines below make the double-hat symbol

\newcommand{\qed}{\nobreak \ifvmode \relax \else
      \ifdim\lastskip<1.5em \hskip-\lastskip
      \hskip1.5em plus0em minus0.5em \fi \nobreak
      \vrule height0.75em width0.5em depth0.25em\fi}

\usepackage{dsfont}

\begin{document}

\title{Dynamical Phase Error in Interacting Topological Quantum Memories}
\author{L. Coopmans\href{https://orcid.org/0000-0001-6501-5420}{\includegraphics[scale=0.066]{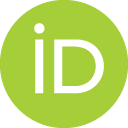}}} 
\email{coopmanl@tcd.ie}
\affiliation{Dublin Institute for Advanced Studies, School of Theoretical Physics, 10 Burlington Rd, Dublin 4, Ireland.}
\affiliation{School of Physics, Trinity College Dublin, College Green, Dublin 2, Ireland}
\author{S. Dooley\href{https://orcid.org/0000-0002-2856-8840}{\includegraphics[scale=0.066]{./ORCID.png}}}
\affiliation{Dublin Institute for Advanced Studies, School of Theoretical Physics, 10 Burlington Rd, Dublin 4, Ireland.}
\author{I. Jubb\href{https://orcid.org/0000-0001-7339-2058}{\includegraphics[scale=0.066]{./ORCID.png}}}
\affiliation{Dublin Institute for Advanced Studies, School of Theoretical Physics, 10 Burlington Rd, Dublin 4, Ireland.}
\author{K. Kavanagh\href{https://orcid.org/0000-0002-6046-8495}{\includegraphics[scale=0.066]{./ORCID.png}}}
\affiliation{Dublin Institute for Advanced Studies, School of Theoretical Physics, 10 Burlington Rd, Dublin 4, Ireland.}
\affiliation{Department of Theoretical Physics, Maynooth University, Maynooth, Co. Kildare, Ireland.}
\author{G. Kells\href{https://orcid.org/0000-0003-3008-8691}{\includegraphics[scale=0.066]{./ORCID.png}}}
\affiliation{Dublin Institute for Advanced Studies, School of Theoretical Physics, 10 Burlington Rd, Dublin 4, Ireland.}
\date{\today}
\begin{abstract}
A local Hamiltonian with Topological Quantum Order (TQO) has a robust ground state degeneracy that makes it an excellent quantum memory candidate. This memory can be corrupted however if part of the state leaves the protected ground state manifold and returns later with a dynamically accrued phase error.  Here we analyse how TQO suppresses this process and use this to quantify the degree to which spectral densities in different topological sectors are correlated. We provide numerical verification of our results by modelling an interacting p-wave superconducting wire. 
\end{abstract}

\preprint{DIAS-STP-21-07}
\maketitle

\section{Introduction}

Topological schemes to protect and manipulate quantum information are based on fractional excitations called anyons. In these approaches information is stored in anyon pairs and, by moving them apart, one can encode this information in a non-local way \cite{Kitaev2001,Dennis2002,Kitaev2003,Kitaev2006,NayakRev,Stanescu2013}. This is the key feature that allows topological memories to be robust against local noise and decoherence processes. 

\begin{figure}
    \centering
    \includegraphics[width=1\linewidth]{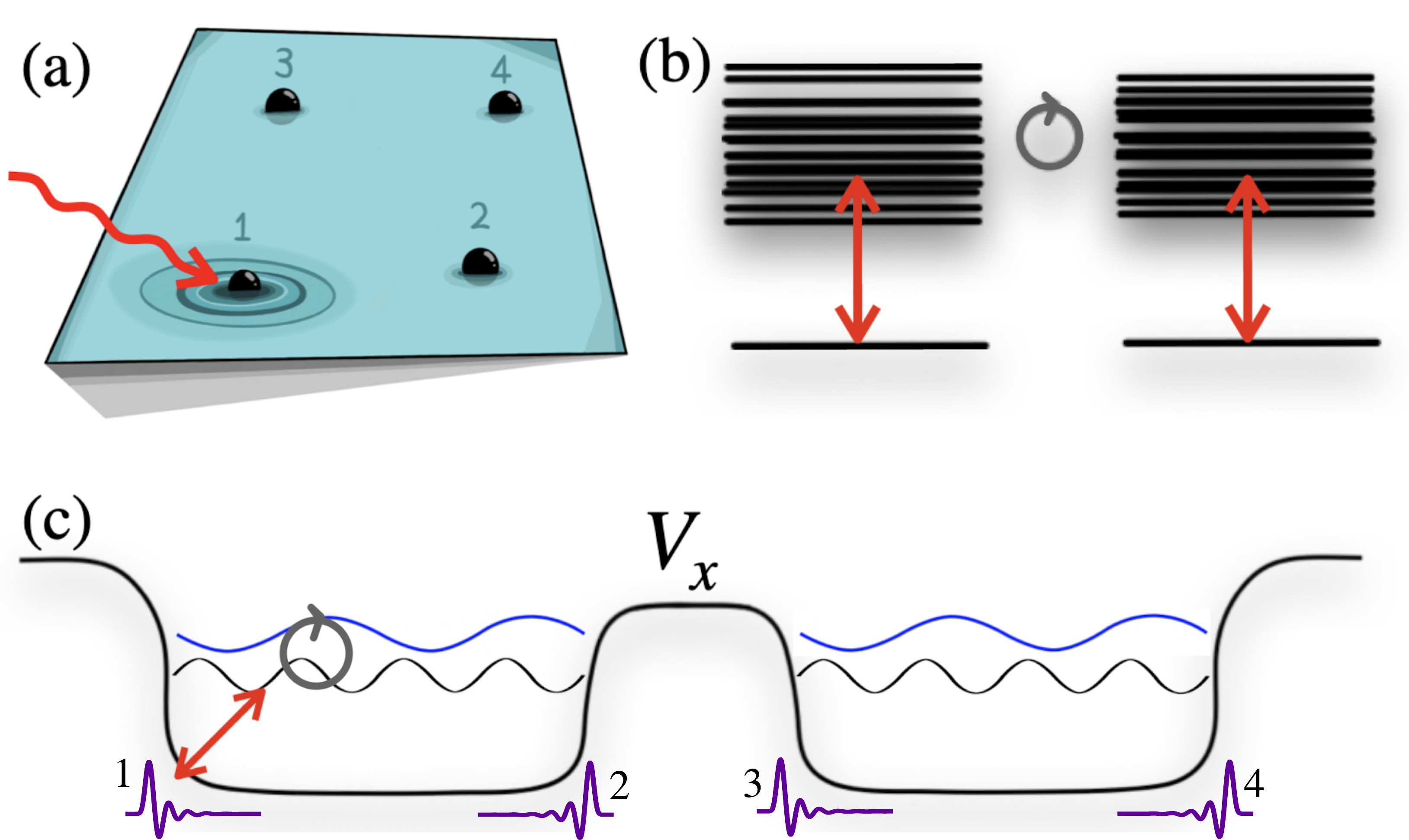}
    \caption{(a) Schematic view of a 2D topological memory consisting of 4 anyons. Information is stored non-locally, and any local excitation occurring near one anyon must propagate through the system for an error to occur. (b) Slight energy mismatches between higher energy states in different topological sectors open up an apparent relative phase error loophole when a local process couples the protected ground states to excited states. (c) As a concrete model we consider a symmetry-protected topological quantum memory consisting of two p-wave superconducting wires separated by a potential barrier. In this setup 4 Majorana zero-modes occur at the domain boundaries, as indicated by the purple lines numbered 1-4 at the bottom. Local fluctuating noise can be realized, for example, by oscillating the left domain boundary, or as a quasi-particle tunneling into the system.}
    \label{fig:Wiresetup}
\end{figure}

In these topological platforms the computational space is a degenerate ground state manifold that emerges when these quasi-particle excitations are far apart (see Fig. \ref{fig:Wiresetup}). The degeneracy of this subspace is fundamentally important because it protects against quantum memory corruption in the form of unwanted qubit rotations, which arise via the accumulation of relative dynamical phases. This feature of the ground state manifold arises from a more general property called Topological Quantum Order (TQO) \cite{PhysRevB.41.9377, Wen:1989iv, Hastings2005, Bravyi2010h, zeng2019quantum}. Among other things, TQO implies that, for states within this manifold, the expectation values of local observables are equal up to some exponentially small correction that depends on the spatial separation between anyons.

Another mechanism which can corrupt the quantum memory with undesirable qubit rotations is dynamical \emph{phase error}. This can occur when the quantum state partially leaves the ground state manifold and returns later with a dynamically altered phase. This type of error could arise on a mean-field level if there are causally connected perturbations near different anyons \cite{Conlon2019}. In this scenario the distance between anyons plays an important role in delaying/reducing the onset of this type of secondary error process. In interacting systems however, a different scenario for phase error still exists because the aforementioned ground state degeneracy does not necessarily extend to a degeneracy in bulk eigenstates. One could then reasonably worry about processes whereby a single localised perturbation creates, and then at a later time, returns an excitation that has accumulated some relative dynamical phase from its time spent in the energy mismatched bulk.

In this paper we show that, despite a mismatched bulk spectrum, TQO guarantees that such interaction driven phase errors are suppressed up to a time that scales with system size. A consequence of this suppression is that, although the bulk spectra from distinct topological sectors differ, there is necessarily a high degree of correlation between spectral densities. This correlation can be quantified by considering appropriate Green’s functions.% involving the ground state manifold. 

We demonstrate these general results, for any system with TQO, using a concrete example: a Majorana based topological memory based on interacting p-wave superconducting wires \cite{Fu2008,Lutchyn2010,Oreg2010,Sau2011,AliceaNat2011}, as shown schematically in Fig.~\ref{fig:Wiresetup}~(c). In this setting we back up our key claims using a TDVP-MPS approach \cite{HaegemanTDVP, Paeckel2019} which allows us to: (1) show that, even in regimes where interactions and time-dependent perturbations are relatively strong, there is no phase error other than what one would expect from mean-field like effects, and (2) directly calculate spectral densities in each topological sector and quantitatively demonstrate the degree of correlation that must exist between them. 

\section{No phase error argument}

Consider some gapped local Hamiltonian~\footnote{By \emph{local Hamiltonian} we mean a sum of Hermitian operators that are each only non-trivial on several neighbouring sites.} $H$ with anyonic quasi-particles. For simplicity, we assume there are two ground states, $\ket{e}$ and $\ket{o}$, corresponding to (e)ven and (o)dd parity respectively~\footnote{This is the case for the Majorana example below. It is straightforward to generalise the argument to a higher dimensional ground state manifold.}. Crucially, we assume that for large enough system diameter $L$, there is a length $L^*$ that scales in some way with $L$ (this length must satisfy $L^*\geq c L$ for some constant $c>0$), for which the ground states satisfy the TQO property: 
\vspace{1mm}

\noindent \emph{For every local operator, $O$, supported in a region of diameter at most $L^*$, we have}
\begin{equation}\label{tqo_assumption}
\bra{e}O\ket{e} = \bra{o}O\ket{o} + \mathcal{O}(e^{-L /\xi})\;\;\; ,
\end{equation}
\emph{for some constant $\xi > 0 $. That is, local operators cannot be used to distinguish the two sectors, up to exponential corrections in $L$.}  
\vspace{1mm}

From the TQO property one can see that the ground states are degenerate, up to exponential corrections in the system length $L$ \cite{Bravyi2010h,Bravyi2011h}. That is, $\bra{a}H\ket{a}$ is approximately the same for $a=e,o$~\footnote{$H$ is a sum of a polynomial (in $L$) number of $k$-local terms, $H_{x}$. TQO implies that {$ \langle a|H_{x} |a \rangle$} is sector independent up to $\mathcal{O}(e^{-L/\xi })$, which dominates the polynomial in $L$ in the sum {$\langle a| H |a \rangle = \sum_x \langle a|H_{x}|a \rangle$}.}.  Although TQO makes no {\em direct} claims on the behaviour of states above the gap that separates the ground state manifold from the bulk excited states, it can be used however to {\em derive} certain properties see e.g.~\cite{Bravyi2010h}. In what follows we show how TQO bounds the dynamically driven phase error, and how this results in a large number of constraints on the bulk spectrum. 

We consider two instantaneous local perturbations of the system, separated by a time $t$. The unitary evolution operator in this case is:
$
U(t) = e^{i \delta'} e^{-i H t} e^{-i \delta}\label{eq:two_kick_unitary} \; ,
$
where $\delta$ and $\delta'$ are local Hermitian operators (possibly at different locations) that do not mix the even and odd sectors. By a \emph{phase error} we mean that $\bra{a}U(t)\ket{a}$ is different for the even and odd sectors. We can expand in the energy eigenstates, $\ket{a ,n}$, to get
\begin{equation}\label{eq:eigenstate_expansion}
\bra{a}U(t)\ket{a} = \sum_n e^{-i E_{a,n} t} \bra{a}e^{i \delta'}\ket{a,n}\bra{a,n} e^{-i \delta}\ket{a},
\end{equation}
where $E_{a,n}$ is the energy of the $n^{th}$ eigenstate in the $a$-sector. For special cases, where there is a degeneracy between the two sectors for all bulk energies, i.e. $E_{e,n}=E_{o,n}$, and where the overlaps $\bra{a,n} e^{-i \delta}\ket{a}$ are sector independent, there is no phase error. Neither condition necessarily holds for an interacting system. Nevertheless, one can make the following argument for the suppression of phase error.

First, we write
\be
\bra{a}U(t)\ket{a}= e^{-i E_0 t}\bra{a} e^{i\delta '}  O(t) \ket{a},
\ee
where we have defined $O(t) := e^{-i H t} e^{-i \delta}e^{i H t}$, and we have used $e^{-i H t} \ket{a} = e^{-i E_0 t} \ket{a}$ where, by the approximate ground state degeneracy, $E_0$ is  exponentially close between the sectors. From the Lieb-Robinson bounds \cite{Lieb:1972wy} we note that we can approximate $O(t)$ by an operator $\tilde{O}(t)$, supported in a region of size $\sim v |t|$ about $\delta$, up to corrections of order $\mathcal{O}(e^{-(L - v |t|)/ \xi})$~\footnote{Our argument involves several exponential error terms; each with its own rate constant $\xi$. To avoid ambiguity, we take $\xi$ to be the smallest such constant such that each error estimate is valid.}. Here $v$ is the Lieb-Robinson speed of propagation for $H$. We can then pick some $T^* \sim L^*/v$, such that the maximum size of $\tilde{O}(t)$ over the range of times $|t| < T^*$ is at most $L^*$, and such that $\tilde{O}(t)$ approximates $O(t)$ to order $\mathcal{O}(e^{-L/\xi})$. There are two cases to consider:

 \begin{figure}
    \centering
    \includegraphics[width=1\linewidth]{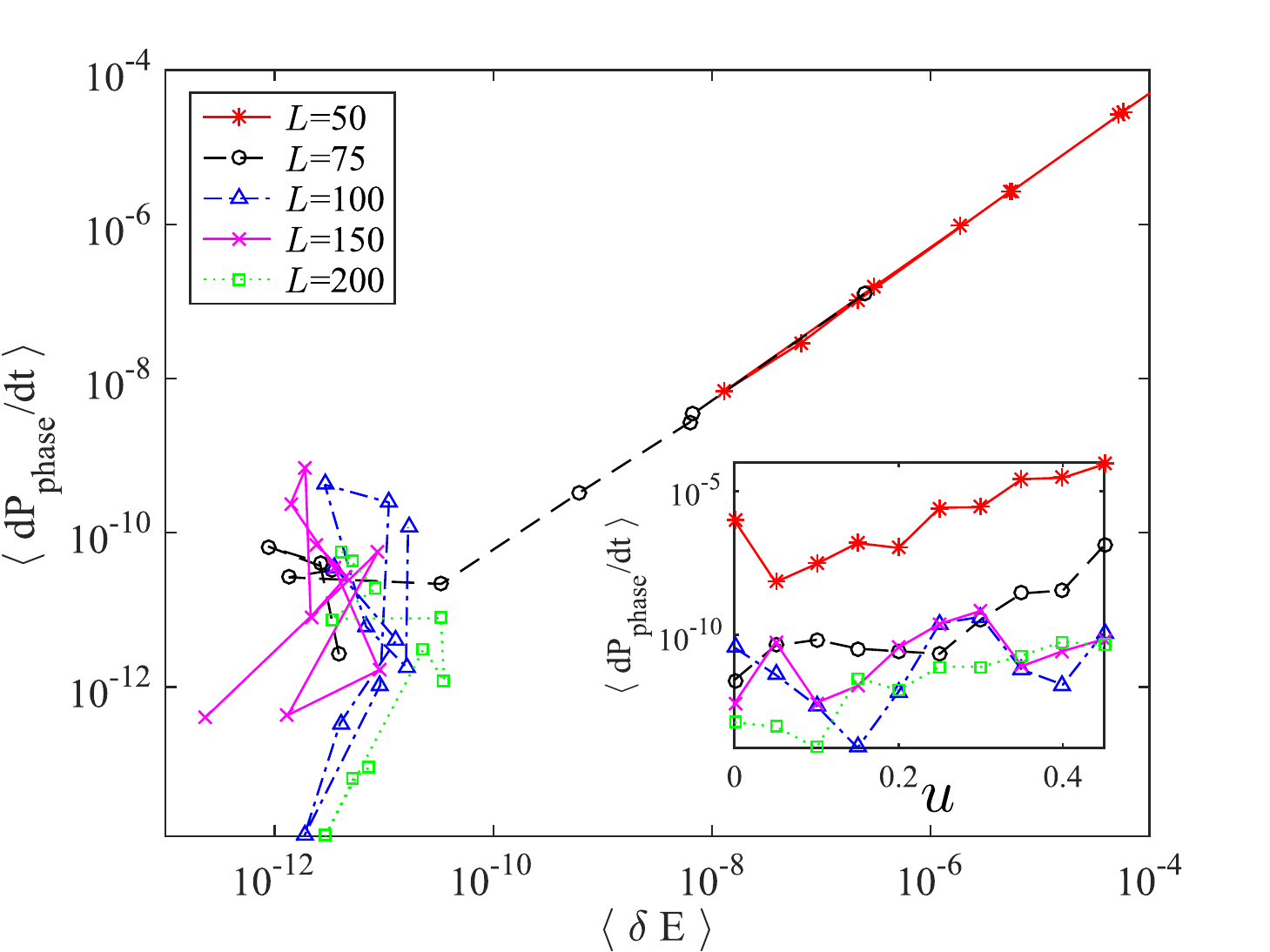}
    \caption{Time-averaged rate of phase error as a function of interaction driven ground state splitting. TDVP can pick up an interaction driven phase rotation (error) resulting from finite-size ground state splitting of order $\langle \delta E \rangle \sim 10^{-10}$. For system sizes where no appreciable splitting occurs we do not detect any systematic phase rotation. Inset: same data as in the main figure plotted against interaction strength (in units of $w$). In these simulations we oscillated the left boundary wall, see App.~\ref{app:MajoranaDetail}, with frequency $\omega = 1$ and maximum velocity $v_{\text{max}} = 0.1$. The time-averaging was done over an $\mathcal{O}(1)$ multiple of the oscillation period. We also set uniform parameters $\mu = -1.5$ and $\Delta = 0.7$. }
    \label{fig:phase_dmrg_oscillating}
\end{figure}
\vspace{1mm}
\noindent \textit{i)} If $\delta'$ is contained in an $L^*$ sized region about $\delta$ for sufficiently large $L$, then, for all times $|t| < T^*$, $e^{i\delta'}\tilde{O}(t)$ is contained in a region of size at most $L^*$. TQO then implies that $\bra{a}e^{i\delta'}\tilde{O}(t)\ket{a}$ is sector independent to order $\mathcal{O}(e^{-L/\xi})$. Putting everything together, for times $|t| < T^*\sim L^*/v$ we find that $\bra{a}U(t)\ket{a}$ is approximately equal between the sectors, up to exponential corrections in $L$. In this situation we say there is \emph{no phase error}.
\vspace{1mm}

\noindent \textit{ii)} If $\delta'$ is not contained in an $L^*$ sized region about $\delta$, then the separation between $\delta$ and $\delta'$ must be growing with $L$. The fact that the system is gapped means that $\bra{a} e^{i\delta'}  \tilde{O}(t) \ket{a} \approx \bra{a} e^{i\delta'} \ket{a}\bra{a}  \tilde{O}(t) \ket{a}$, up to an exponential correction in the operator separation~\citep{Hastings_2004,Hastings:2005pr}, and hence of order $\mathcal{O}(e^{-L/\xi})$. For $|t| < T^*$ the TQO condition implies that $\bra{a}  \tilde{O}(t) \ket{a}$ is sector independent up to exponential corrections, and hence $\bra{a}e^{i\delta'}  \tilde{O}(t) \ket{a}$ is too. That is, we again have exponentially vanishing phase error for times $|t| < T^*$. 
\vspace{1mm}

 \begin{figure}
    \centering
    \includegraphics[width=1\linewidth]{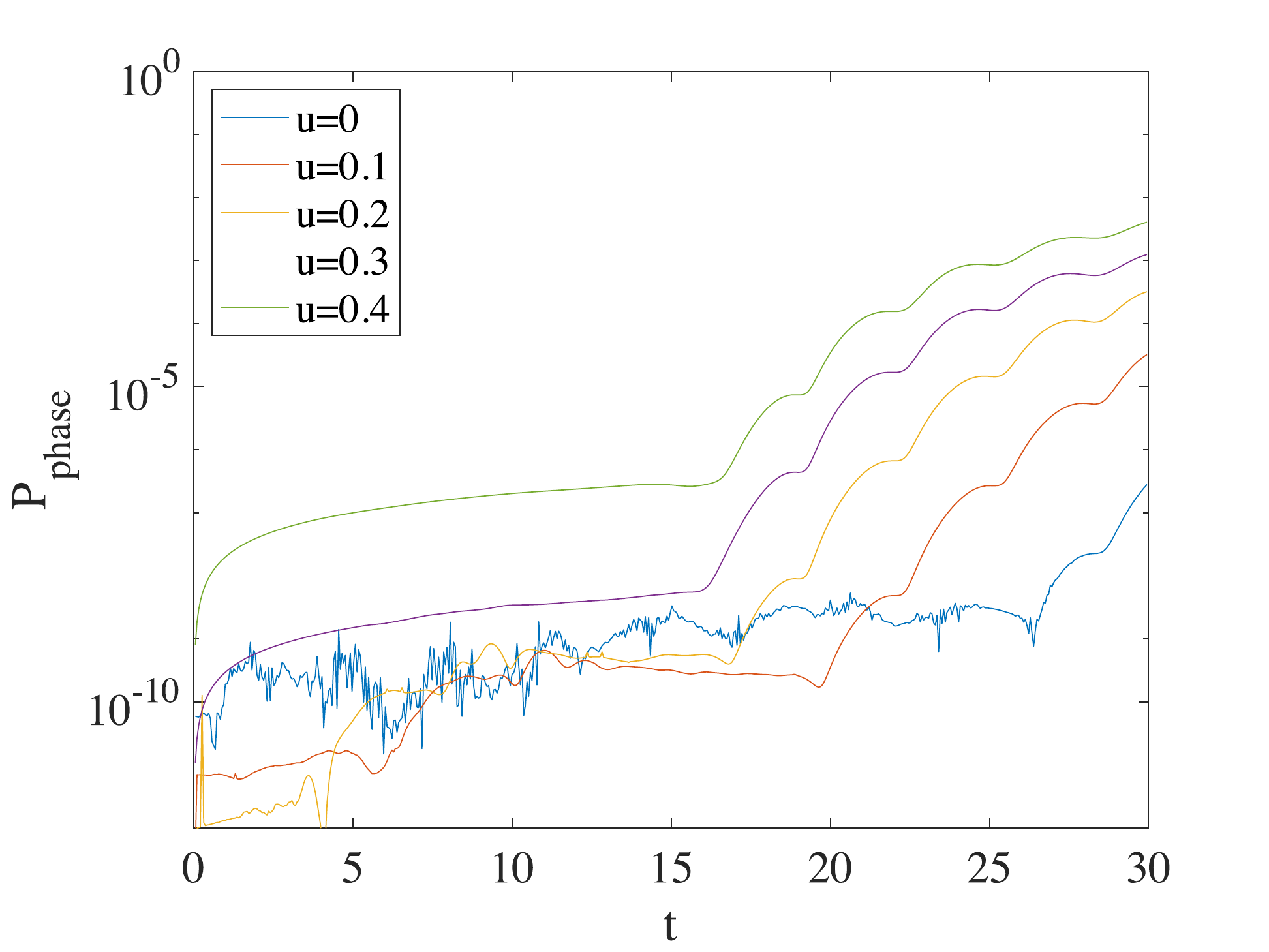}
    \caption{Phase error for two oscillating walls. In this simulation we set uniform parameters $w = 1$, $\Delta = 0.8$, $\mu = -1.2$, $L = 70$,  $dt= \pi/60$ and a bond dimension of $\chi = 50$.  Both walls were oscillated with frequency $\omega = 1$ and  maximum velocity $v_{\text{max}} = 0.1$. }
    \label{fig:phase_dmrg_oscillating2}
\end{figure}

Note that the exact time, $T^*$, for which no phase error arises, depends upon the specific system, the perturbations, and their separation. The preceding argument should be understood as heuristic justification that $T^*$ scales as $L^*/v$, and hence as $L/v$. In Sec.~\ref{app:TimeDepPert} we address the general scenario of time-dependent perturbations, and we make similar arguments for phase error suppression.

The suppression of phase error implies a large number of constraints on the bulk spectra in the following way. From~\eqref{eq:eigenstate_expansion} we can write the amplitude of the phase error, denoted here by $\alpha(t)$, as:
\begin{align}\label{eq:spectral_diff_bound}
\alpha(t) & = \bra{e} U(t) \ket{e} -  \bra{o} U(t) \ket{o} 
\\
& = \sum_{n>0}  \lambda_{e,n,\delta \delta'}e^{-i E_{e,n} t} -  \lambda_{o,n,\delta \delta'} e^{-i E_{o,n} t} + \mathcal{O}(e^{-L / \xi })  \; \nonumber ,
\end{align}
where $\lambda_{a,n,\delta \delta'} = \bra{a}e^{i \delta'}\ket{a,n}\bra{a,n} e^{-i \delta}\ket{a}$, and the sum runs from $n>0$ as the ground state contribution is contained in the exponential correction on the far RHS. Above we showed that $| \alpha (t) |$ is exponentially suppressed. This then implies that the magnitude of difference between the even and the odd sums (for $n>0$) on the RHS is also exponentially suppressed in $L$. Crucially, this is also the case for many pairs of local operators, $\delta$ and $\delta'$, though the exact form of the suppression may differ in each case. One can then consider this constraint between the even and odd sums for \emph{all} such pairs, resulting in a substantial number of constraints. The effect of this on the bulk splittings is illustrated in Sec.~\ref{sec:MajoranaEx} for a specific model.

\subsection{No Phase Error for Time-Dependent Perturbations}\label{app:TimeDepPert}

Here we generalize the no phase error argument to time-dependent perturbations. Consider a time-dependent local perturbation $\delta(t)$. The time-ordered unitary evolution operator is then 
\be
U(t) = T \{ e^{-i \int_0^t dt' H + \delta(t') } \} \; .
\ee
To show there is no phase error we need to show that $\bra{a}U(t)\ket{a}$ is approximately sector independent. As $U(t)$ is not a local operator we cannot immediately use the TQO condition. However, we can turn it into a local operator times a non-local operator that acts trivially on the system eigenstates:
\begin{equation}\label{eq:interaction_picture_unitary}
U(t) = e^{-i H t} \times T \{ e^{-i \int_0^t dt' \tilde{\delta}(t') } \},
\end{equation}
where $\tilde{\delta}(t):= e^{i H t}\delta(t)e^{-i H t}$, which essentially follows from the interaction picture.  

From the Lieb-Robinson bounds, we can approximate $\tilde{\delta}(t)$, and hence the time-ordered unitary involving $\tilde{\delta}(t)$, by an operator supported in a region of size $\sim v |t|$, up to corrections of order $\mathcal{O}(e^{-(L- v|t|)/\xi})$. Thus, we have
\begin{align}
\bra{a}U(t)\ket{a} & = \bra{a}e^{-i H t} \times T \{ e^{-i \int_0^t dt' \tilde{\delta}(t') } \}\ket{a} \nonumber
\\
& = e^{-i E_0 t} \bra{a} T \{ e^{-i \int_0^t dt' \tilde{\delta}(t') } \}\ket{a} \;\;\;,
\end{align}
where $E_0$ is the same in both sectors up to corrections of order $\mathcal{O}(e^{-L/\xi})$. For $|t| < T^* \sim L^*/v$, we make an error of order $\mathcal{O}(e^{-L/\xi})$ by approximating the time-ordered unitary in a region of size $L^*$. In doing so we can then use TQO to argue that the expectation value on the last line is sector independent to order $\mathcal{O}(e^{-L/\xi})$, and hence $\bra{a}U(t)\ket{a}$ is too, i.e. no phase error.

If we now include a second time-dependent perturbation $\delta'(t)$, separated from $\delta(t)$ such that $[\tilde{\delta}(t),\tilde{\delta}'(t)]\approx 0$ up to exponential corrections in the separation between $\delta (t)$ and $\delta' (t)$, then the unitary $U(t)$ approximately factorises as
\begin{equation}
U(t) \approx e^{-i H t} \times T \{ e^{-i \int_0^t dt' \tilde{\delta}(t') } \} \times T \{ e^{-i \int_0^t dt' \tilde{\delta}'(t') } \} .
\end{equation}
It is straightforward to apply the gap argument from above to show that this will also not incur a phase error for $|t|<T^* \sim L^*/v$.

\section{A Majorana-based Example}\label{sec:MajoranaEx}
To demonstrate these general results we model a simple topological memory consisting of two one-dimensional p-wave superconducting wires separated by a potential barrier, as shown in Fig. \ref{fig:Wiresetup} (c). The $2L$-site ($L$ sites each for the left and right wires) lattice Hamiltonian \cite{Kitaev2001} for this is given by:
\begin{equation}
\label{eq:Hamiltonian}
\begin{split}
	H_0 &=  - \sum_{x=1}^{2L} [\mu_x-V_x]  (c^\dagger_{x}c_{x }-1/2) \\
	-&  \sum_{x=1}^{2L} (w \, c^\dagger_{x}c_{x+1} + h.c) + \sum_{x=1}^{2L}(\Delta_x c^\dagger_{x} c^\dagger_{x+1} + h.c. ),
\end{split}
\end{equation}
where $\Delta_x$ is the superconducting gap, $\mu_x$ is the on-site chemical potential, $V_x$ is the potential profile, see App.~\ref{app:MajoranaDetail}, $w$ is the hopping parameter and the $c^{(\dagger)}_x$ represent fermion (creation) annihilation operators. Interactions are modelled using a nearest-neighbour density-density term:
	\begin{equation}\label{eq:Hint}
	H_{\text{int}}= \sum_{x=1}^{2L-1} u_x c^\dagger_x c^\phd_x c^\dagger_{x+1} c^\phd_{x+1},\end{equation}
where $u_x$ is the interaction strength. The fully interacting Hamiltonian of interest is $H = H_0 + H_{\text{int}}$. This model can be effectively realized in proximity coupled systems \cite{Fu2008,Lutchyn2010,Oreg2010,Sau2011,AliceaNat2011,Stanescu2013} which are backed up with extensive experimental evidence \cite{Kouwenhoven2012,Deng2012,
Das2012,Finck2013,Churchill2013,
Albrecht2016,Zhang2016,Deng2016,
NadjPerge2014,Ruby2015,Pawlak2016,
Zhang2018,Marcus2019,Falko2019}.  For discussions on different types of noise that can occur in these systems see~\cite{Rainis2012,Budich2012,Roy2012,Konschelle2013,Yang2014,Ng2015,Pedrocchi2015,Hu2015,Ippoliti2016,Brown2016,Aseev2018,Knapp2018,Zhang2019,Karzig2013,Scheurer2013,Conlon2019,coopmans2020protocol}. 

The non-interacting Hamiltonian, $H_0$, gives rise to four Majorana zero-energy modes $\Gamma_j$ \cite{Kitaev2001}, localised at the domain walls between topological ($\mu_x+V_x \leq 2w$) and non-topological regions encoded by $V_x$, which can be paired into two Dirac fermionic zero-modes $\hat{\beta}_0^{\mathtt{L}/\mathtt{R}}=\frac{1}{\sqrt{2}}(\Gamma_{1/3}+i\Gamma_{2/4}$), one for the left ($\mathtt{L}$) and one for the right wire ($\mathtt{R}$). From the 4-fold degeneracy associated with these zero-modes we can, without loss of generality, define our topologically protected ground states from the even-parity sector (states with an even number of excitations) as:
\begin{equation}
\label{eq:groundstate}
\begin{split}
\ket{ 00 \{0 \}} = \ket{0}_{\mathtt{L}}  \ket{0}_{\mathtt{R}} \ket{\{0 \}} \quad
\ket{ 11 \{0 \}}= \ket{1}_{\mathtt{L}} \ket{1}_{\mathtt{R}}  \ket{\{0 \}},
\end{split}
\end{equation}
in which $\ket{\{0 \}}= \ket{0}^{\otimes L-2}$ corresponds to the ground state of the bulk modes and $\ket{0}_\alpha,\ket{1}_\alpha$ are the respective unoccupied and occupied Dirac zero-mode for the $\alpha=\mathtt{L}/\mathtt{R}$ wire. In what follows we will effectively forget about the right wire. Its purpose is to keep us in the even-parity sector overall. Henceforth, any mention of even and odd sectors, and wires, will refer to the left wire only.

When local interactions are present ($u_x\neq 0$) we can no longer label many-body eigenstates of $H$ in terms of their quasi-particle occupation numbers. However, it can be shown that the topological ground state degeneracy holds (up to exponential corrections in the length of the system) provided that the interactions do not close the gap \cite{Moon2018}, and that the system satisfies the TQO property.
%When local interactions are present ($u_x\neq 0$) the topological {\em ground state} degeneracy holds (up to exponential corrections in the length of the system). 
However, the same degeneracy between the excited states can be broken by interactions when the spectrum contains overlapping quasi-particle bands~\cite{Kells2015,Moran2017,Kells2018b,Pellegrino2020}, leaving open the possibility for  phase errors to be returned to the ground state without the need for quasi-particle propagation. The results presented here are focused solely upon such dynamically generated phase errors due to bulk energy mismatches. However, based on our argument in the previous section, this mechanism must be suppressed because $H$ has TQO as shown in App.~\ref{app:TQOinher}.  %Numerical confirmation of this is presented below.

\subsection{Numerical Verification} To support our analytical argument we now present numerical results for the more general case of a time-dependent perturbation. Specifically, we simulate an oscillating perturbation $\delta(t)$ on the left boundary of the Majorana wire, see App.~\ref{app:MajoranaDetail}. We quantify the phase error as \begin{equation}P_{\text{phase}}(t) := \frac{1}{4}| \bra{e}U(t)\ket{e} - \bra{o}U(t)\ket{o} |^2
\end{equation} in what follows. Note that $P_{\text{phase}}(t)=0$ corresponds to no phase error. 

Our numerical results are obtained using time-dependent-variational-principle (TVDP) matrix product state (MPS) simulations~\cite{HaegemanTDVP,Paeckel2019}. In Fig. \ref{fig:phase_dmrg_oscillating}, we examine the time-averaged rate of phase error as a function of the the time-averaged energy difference, $\langle \delta E \rangle$, between the approximately degenerate ground states for finite size systems \footnote{The ground state energy splitting is time-averaged because the spectrum to varying in time due to the time-dependent perturbation.}. The results clearly show that the numerical methodology is capable of detecting phase error in the dynamics once the ground state splitting is of order $\langle \delta E \rangle\sim 10^{-10}$. When the system is large enough, such that interactions do not make a detectable change to $\langle \delta E \rangle$, no systematic increase in phase error occurs.

We also consider the case of a second oscillating perturbation $\delta'(t)$ at the right boundary of the wire. Fig.~\ref{fig:phase_dmrg_oscillating2} shows that in this case, after a time $T^* \sim L/v$ there is a sudden increase in phase error. This is well understood on a mean field level: excitations originating at one wall return to the ground state space at the other wall, generating a $\sigma^z$ error in the ground state space (see, e.g., \cite{Scheurer2013, Conlon2019}). Crucially, before $T^*$ we see no evidence of any other systematic phase error, other than that attributable to the splitting in the ground state manifold.  

In addition to these phase error results we also provide some numerical TDVP-MPS simulation results for the qubit-loss error (infidelity) as defined by
\begin{equation}
P_{\text{loss}}(t) := 1-| \bra{e}U(t)\ket{e}|^2 - |\bra{o}U(t)\ket{o} |^2.
\end{equation}

\begin{figure}[ht!]
    \centering
    \includegraphics[width=1.0\linewidth]{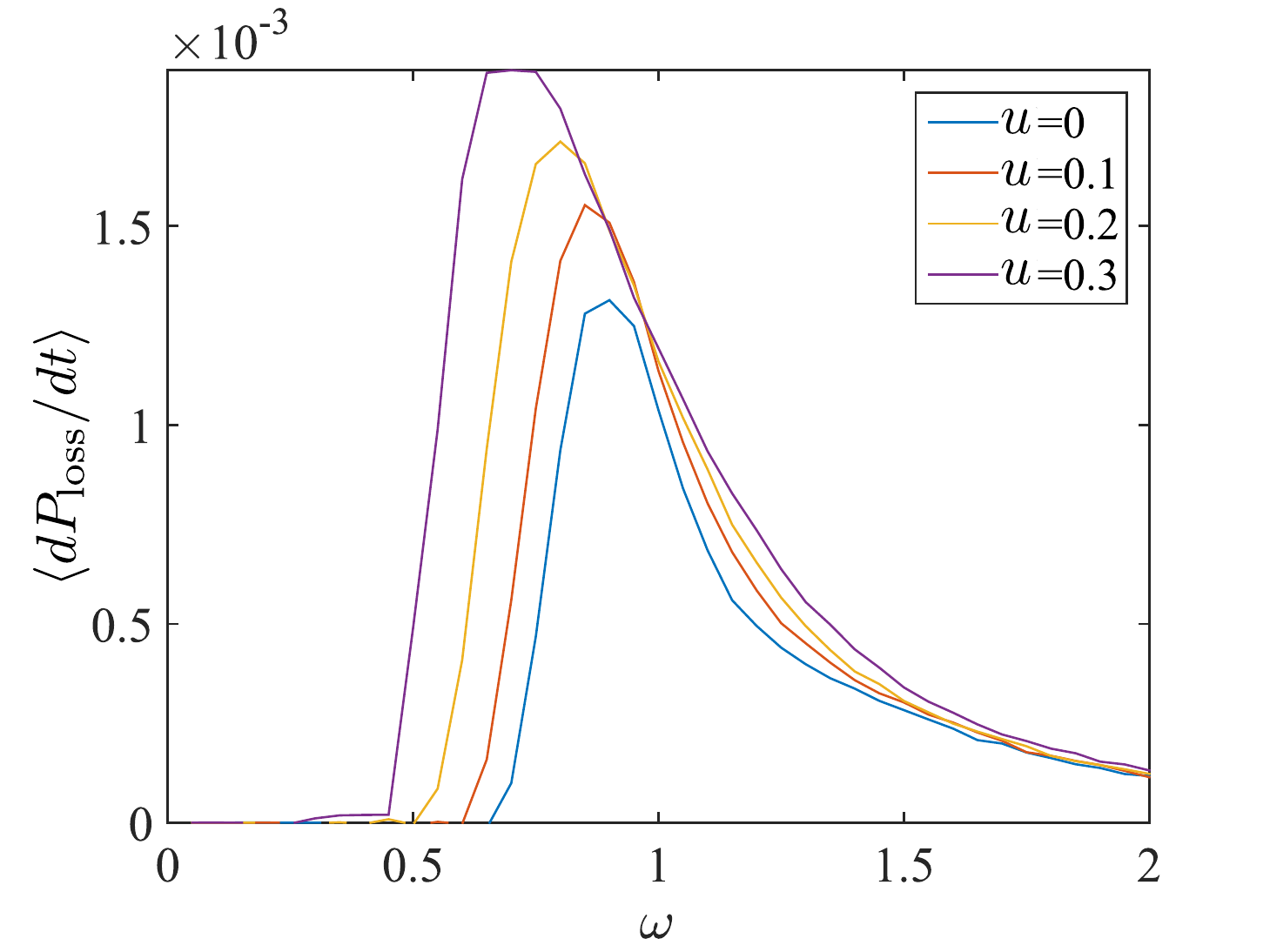}
    \caption{(Color online): Averaged rate of qubit-loss versus frequency of the left boundary wall for 4 different interaction strengths. The frequency covers a wide regime from slow almost adiabatic oscillations, $\omega \ll E_{\text{gap}}$, to high (non-adiabatic) frequencies $\omega \gg E_{\text{gap}}$. The resonance peak shifts to lower frequencies with increasing interaction strength, corresponding to an expected reduction in the gap. Moreover across all frequencies the rate of qubit-loss increases with interaction strength. For this plot we set $L=50$, $v_{\text{max}}=0.1$, $w=1$, $\mu = -1$ and $\Delta = 0.5$. }
   \label{fig:omegaU_Qubitloss}
\end{figure}

In Fig.~\ref{fig:omegaU_Qubitloss} we show the response of the system to a single oscillating boundary across a range of frequencies from the almost adiabatic regime, $\omega \ll E_{\text{gap}}$, up to frequencies far in excess of the topological gap. The results resemble the expected local density of states in the wire, in a given parity sector. We see that repulsive interactions lower the peak resonant frequency (in agreement with an expected reduction to the band edge). We also see that the topological memory degrades slightly across all frequencies. This is somewhat different from what is observed in non-interacting studies, where a lower bulk gap from, say from a reduced p-wave pairing or a lower wire electron density, can result in an improved robustness to high-frequency perturbations \cite{Conlon2019}.

\section{Spectral correlation} Notably, the deviations observed in phase error after $T^*$ are not driven by the phase error accumulated in the bulk states, but are associated with the fact that the dynamics eventually make local operators non-local. However, the fact that we cannot observe phase error before $T^*$, for all initially local perturbations, implies that the even-odd excited energy spectra in the interacting model are related in some way that is unique to systems with TQO.  This is most clearly demonstrated by Fourier transforming the system's Green's functions.

\begin{figure*}
    \centering
    \includegraphics[width=1\linewidth]{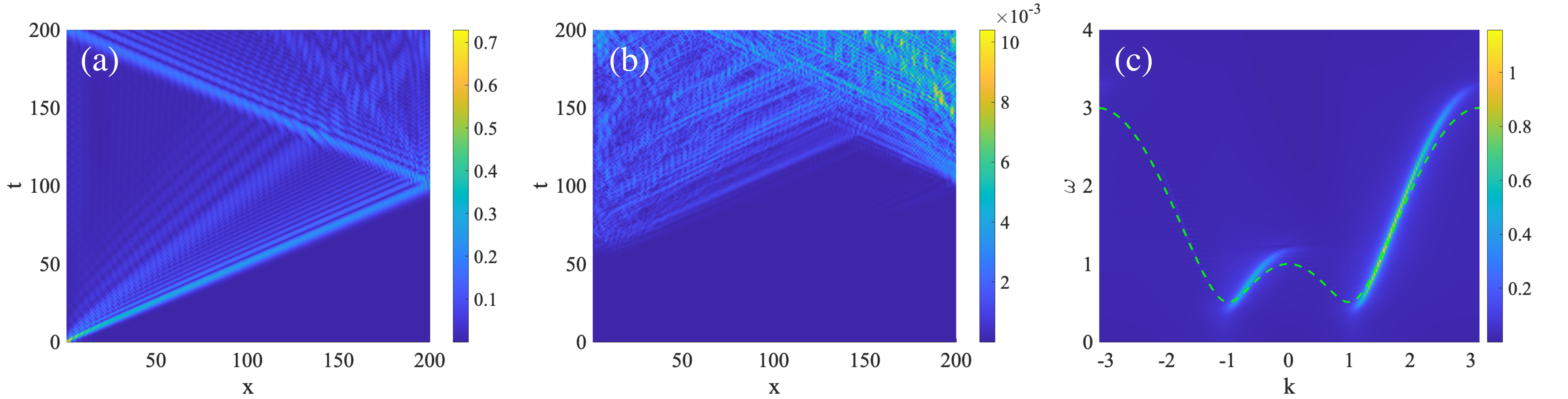}
    \caption{The single particle Green's functions (a) $|G^{e} (1,x,t)|$ and (b) the difference  $|D|=|G^{e}(1,x,t)-G^{o}(1,x,t)|$, which shows that the two correlators are the same up to exponentially small corrections for a time $T^*$. (c) Introducing a cut-off in time, $\Theta (t) =e^{- .03 |t|}$, leads to spectral density $\mathcal{F}[\Theta \times G^a](k,\omega)$ that is weighted around positive group velocities. The dashed line shows the free single particle dispersion for comparison. The resolution of $\mathcal{F}[\Theta \times G^e]$ is set by $1/L$ for momenta and $1/T^* \sim v/L$ for frequency. At this resolution there are only exponentially small differences between $\mathcal{F}[\Theta \times G^e]$ and $\mathcal{F}[\Theta \times G^o]$.}
    \label{fig:GF_res_fermion}
\end{figure*}

Consider for example taking $e^{i \delta'}= 2 c^\dagger_{x'}c_{x'} -\mathds{1}$ and $e^{-i \delta} = 2 c^\dagger_{x} c_{x}-\mathds{1}$ in $U(t)$ %in \eqref{eq:two_kick_unitary} 
and define the Green's function for the $a$-sector as
\begin{equation}
    G^{a}(x,x',t) := e^{i E_0 t } \langle a | U(t) |a \rangle \; ,
\end{equation}
which depends on the operator positions, $x$ and $x'$, and the time $t$. For convenience, we fix $x=1$ (the left boundary of the wire) or $x=L/2$ (the middle of the wire) %in what follows, 
and consider $G^a$ as a function of $(x',t)$ only.

No phase error implies that the difference, $D := G^e - G^o,$ is exponentially suppressed for times $|t| < T^*$, and hence so is the integral of $D$ over $[-T^*, T^*]$. One can then verify that this implies that the Fourier transform, $\mathcal{F}[D](k,\omega)$, is exponentially suppressed when convolved with the Fourier transform, $\mathcal{F}[\Theta]$, of some cut-off or box function, e.g. $\Theta(t) \sim 1$ for $|t| < T^*$ and $\sim 0$ otherwise. In other words, while the spectral functions, $\mathcal{F}[G^a](k,\omega)$, differ between sectors, these differences are washed out by smearing, or convolving, with $\mathcal{F}[\Theta]$~\footnote{For a box function, with cut-off $T^{*}$, $\mathcal{F}[\Theta]$ has the form of a sinc function of width $\sim 1/T^*$, and hence a convolution with $\mathcal{F}[\Theta]$ acts to smear $\mathcal{F}[G^a](k,\omega)$ by this width in the $\omega$ direction.}. It is important to note that this is not a perturbative statement.

\subsection{Extension to fermionic operators} The analytical arguments above rest heavily on the Lieb-Robinson bounds, which are formulated as commutation relations between local operators. Fermionic operators such as $c_x$ are non-local, as they can be seen as strings of local operators through the Jordan-Wigner representation. Even powers of such fermionic operators, on the other hand, are local. Our analytical results, therefore, only pertain to the latter, and hence only address time-dependent changes within parity preserving Hamiltonians. 

Despite this, in Fig.~\ref{fig:GF_res_fermion} we illustrate the Green's functions and their FT's for fermionic operators $e^{i \delta'}  =  c^\dagger_{x'}+ c_{x'}$ and $e^{-i \delta}  =  c^\dagger_{1} + c_{1} $.  The resulting resolution in reciprocal space $[k \times \omega ]$ is of the order $[1/L \times v/L$].  One can estimate from Fig.~\ref{fig:phase_dmrg_oscillating2} that the Lieb-Robinson speed $v\propto u$, and hence, while increased interactions imply a coarser grained energy correlation between sectors, making the system larger allows us to systematically improve this bound.  For details on how similar arguments can apply to multiparticle Green's functions see App.~\ref{app:MultPartGF} for further discussion. The numerical results, in particular the cancellations of even and odd Green's functions up to a time $T^*$ (Fig. \ref{fig:GF_res_fermion} (b)), indicate that it may be possible to extend the analytical argument to fermionic perturbations. The work in~\cite{2017arXiv170508553N,Hastings:2005pr} likely offers a good starting point for such generalisations.  

\section{Conclusion}
We have shown that in systems possessing TQO dynamical driven phase-errors are suppressed which implies that spectral densities in different topological sectors are always highly correlated. This result offers a useful baseline for research into strong zero-modes and high-temperature properties of topological memories. Although the bounds presented are weaker than the special cases of strong zero-modes \cite{Gangadharaiah2011, Jermyn2014, Kells2015, Kells2015b, Fendley2016, Kemp2017, Moran2017, McGinley2017, Ardonne2018, Pellegrino2020, Ardonne2020}, they are both general and non-perturbative, applying to any material that satisfies the TQO condition. This is important, given the growing number of novel materials with topological features~\cite{Bradlyn2017} of which some may satisfy the TQO condition.  

More generally it might be possible to tighten rigorous bounds such as \cite{Bravyi2010h} and make contact with works concerning disorder and constrained thermalisation e.g.  \cite{Huse2013, Else2017, Kells2018b, Nulty2020}.  Although it is well established that disorder prevents stray excitations propagating across a system (see e.g. \cite{Conlon2019}), for free-fermion systems there is no evidence of a reduction in the propagation time itself. The situation could be very different in interacting environments and so a TDVP-MPS examination of this scenario would be an interesting avenue for further study. 

\section*{Acknowledgements}
We would like to acknowledge D. Pellegrino, A. Conlon, J. K. Slingerland, A. Romito, D. Meidan, F. Pientka and A. Werner for fruitful discussions, and A. Kells for her artistic renderings. L.C., K.K., S.D., and G.K. acknowledge Science Foundation Ireland for financial support through Career Development Award 15/CDA/3240. G.K. was also supported by a Schr\"{o}dinger Fellowship. IJ is supported by a Schr\"{o}dinger Scholarship.

\appendix

\section{Potential profile}\label{sec:numerical_setup}\label{app:MajoranaDetail}
We simulate the two-wire setup of the Majorana based topological memory by utilising the potential profile:
\begin{equation}
\label{eq:Potential}
\begin{split}
    V_x &= V_{\text{outer}} [ f(x_1-x)+f(x-x_4)] \\ 
    &+ V_{\text{inner}} [ f(x-x_2)-f_3(x-x_3) ],
\end{split}
\end{equation}
where $x_i$, for $i=1,2,3,4$, encodes the lattice site positions of 4 domain walls, and $f(x)= (1+\exp(-x/\sigma))^{-1}$ is a Sigmoid function. We also choose the separating potential, and separation $|x_{2}-x_{3}|$, to be large enough so as to render both wires fully disconnected. Note that any time-dependent noise protocols above only influence the left wire, and we assume that all low energy bulk excitations occur only in the left wire. 

In Fig. \ref{fig:phase_dmrg_oscillating} and \ref{fig:phase_dmrg_oscillating2} in the main text we oscillate the walls of the left wire. This is implemented by replacing $x_{1,2}$ with a time-dependent function $x_{1,2}(t)=( v_{\text{max}}/\omega ) \sin{\omega t}$, with a velocity $v_{\text{max}}$ and frequency $\omega$ that does not break \newline(super-)adiabaticity~\cite{Karzig2013,Scheurer2013,Conlon2019,
coopmans2020protocol}. 

\section{TQO inheritance}\label{app:TQOinher}

Here we apply the arguments in~\cite{Hastings2005} to show how the interacting Majorana wire, $H$, inherits \emph{approximate} TQO~(Eq.~\ref{tqo_assumption} in the main text) from some simpler system, denoted here by $H'$, with \emph{exact} TQO.  That is, the expectation values of any sufficiently local operator in the even and odd ground states of $H'$ are \emph{exactly} equal. For $H$ to inherit TQO from $H'$ we require the two Hamiltonians to be quasi-adiabatically connected via a local process that does not close the gap.

There are several choices for $H'$ that satisfy these requirements. For particular parameters, $H_0$ (the non-interacting wire~Eq.~\ref{eq:Hamiltonian} in the main text) has exact TQO for any operator of size at most $L-1$~\cite{Kitaev2001}, and one can tune from $H_0$ to $H$ by locally tuning the interaction strength and the other parameters in Eqs.~\ref{eq:Hamiltonian} and \ref{eq:Hint} in the main text in a way that does not close the gap. 

\begin{figure*}
\centering
\subfigure[\,  $\log_{10} |G^e(x,t) -G^o(x,t)|$ ]{
\includegraphics[width=.31\textwidth,height=0.2\textwidth]{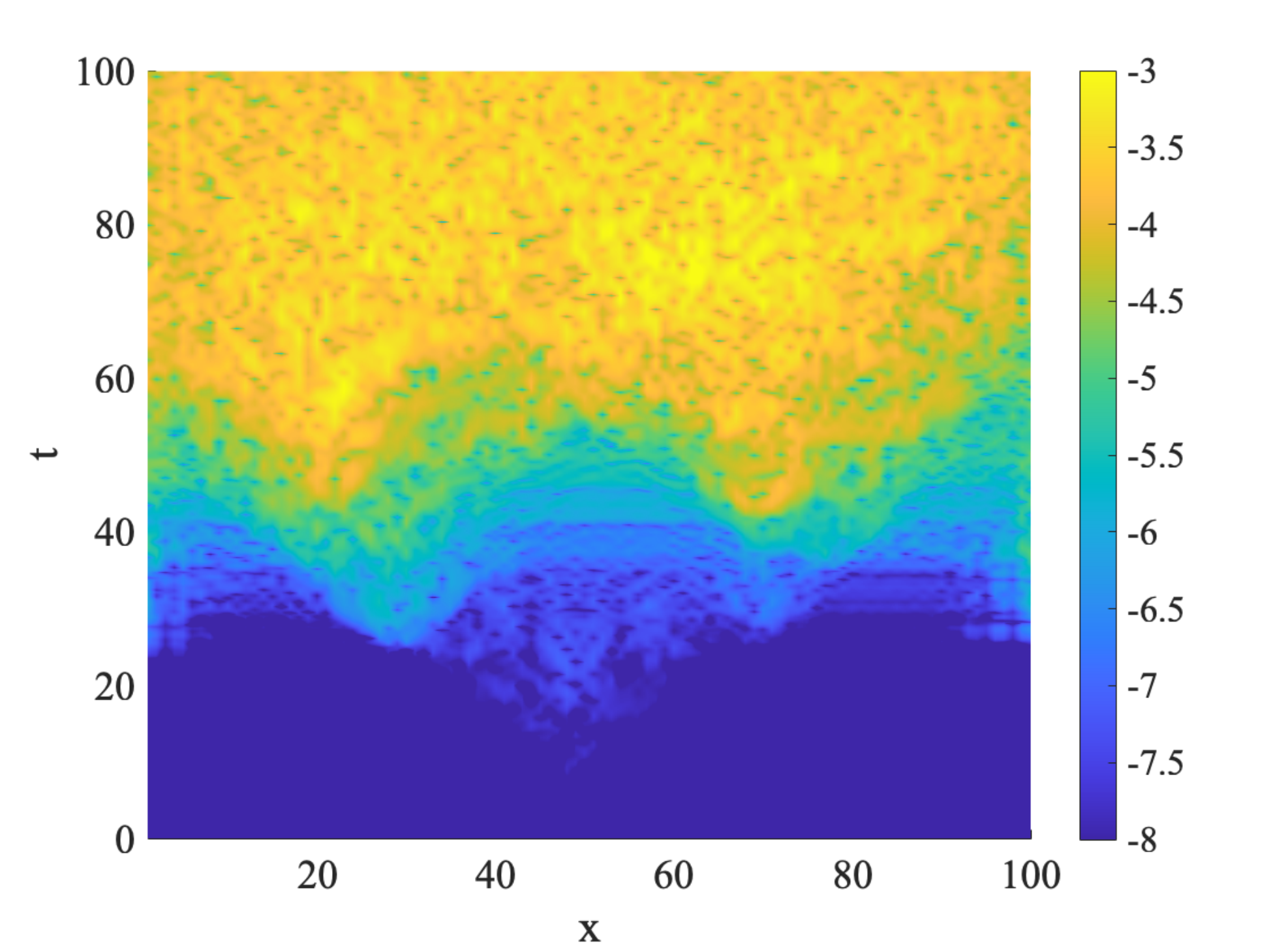}
\label{fig:LD}
}
\subfigure[\,  $ \mathcal{F}(\Theta \times G^e)$  ]{
\includegraphics[width=.31\textwidth,height=0.2\textwidth]{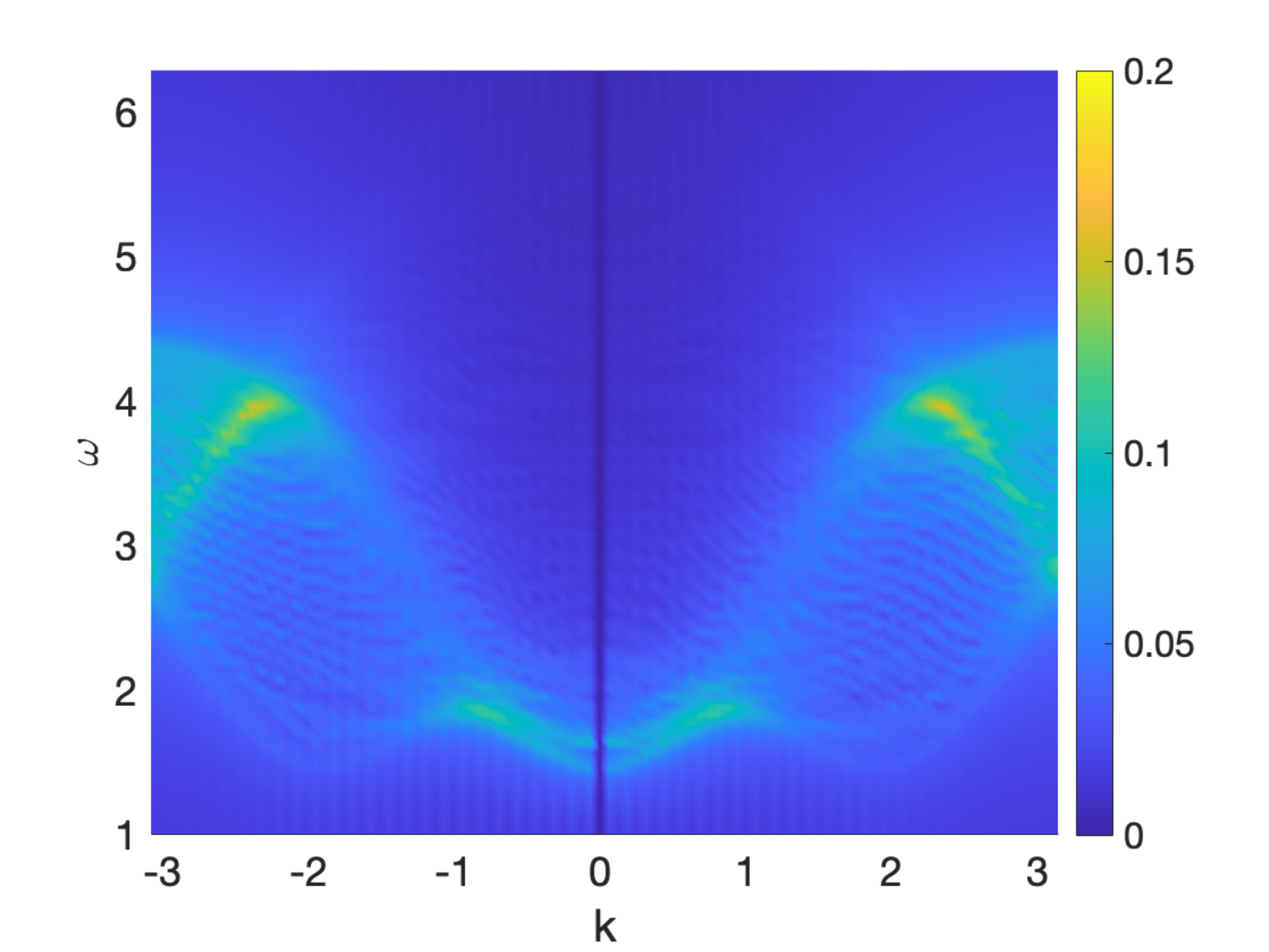}
\label{fig:FTG}
}
\subfigure[\, 2-particle density]{
\includegraphics[width=.31\textwidth,height=0.2\textwidth]{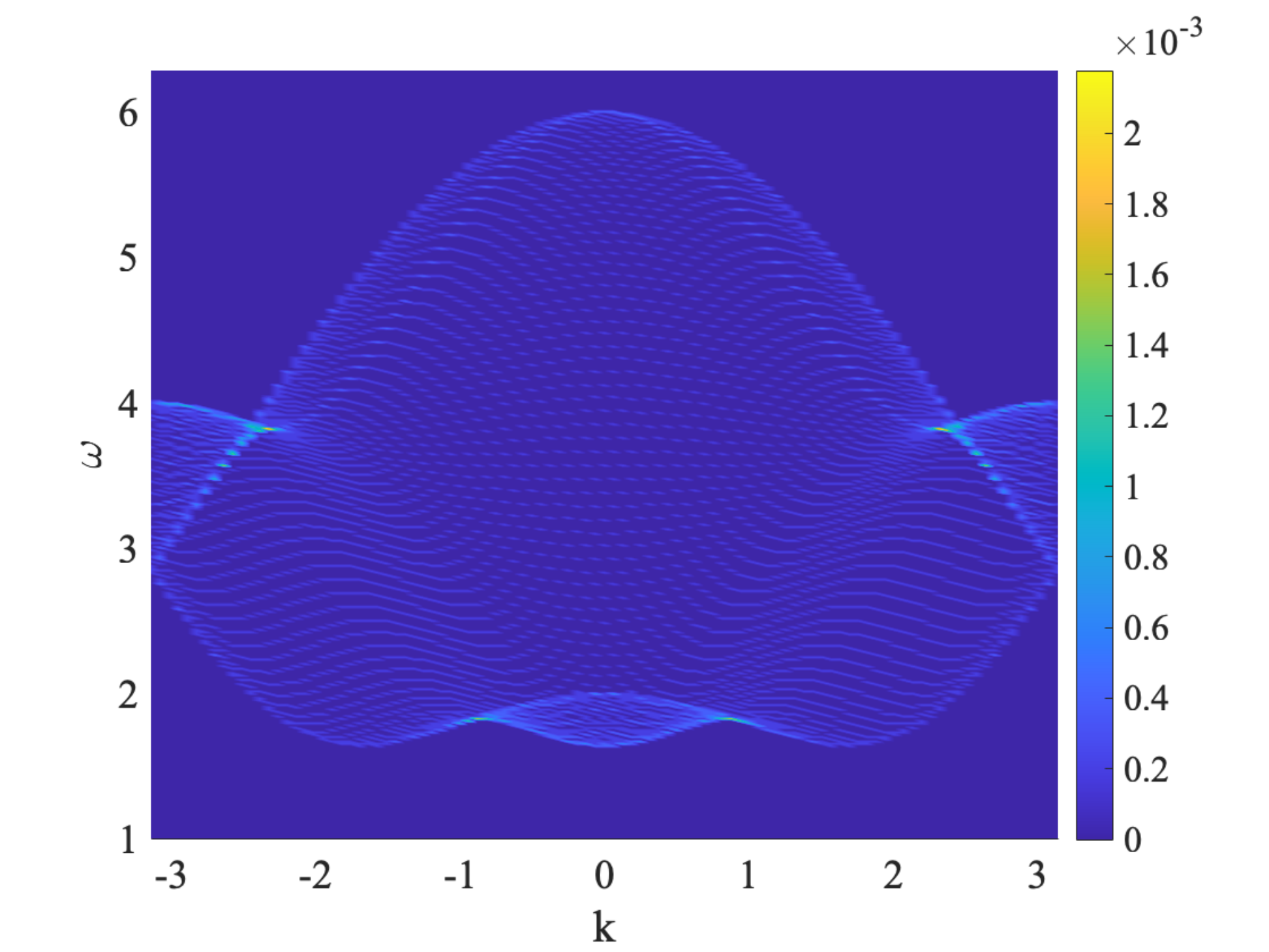}
\label{fig:TPD}
}
\caption[]{(Color online): (a) The difference $|G^e(x,t) -G^o(x,t)|$ is negligible up to a time $t=T^{*}$. (b) As a result, $ \mathcal{F}(\Theta \times G^e)\approx \mathcal{F}(\Theta \times G^o)$ for a cut-off function $\Theta(t)$ that drops off after $T^*$. (c) The total momentum resolved 2-particle spectral density of the non-interacting system ($u_x=0$) for a system size of $L=100$.}
\label{fig:TP_spectral}
\end{figure*}

A more useful choice for $H'$, due to its proximity to $H$ in parameter space, is a partially interacting wire. Specifically, we tune the couplings on one side of the wire (say the right) to what we refer to as the \emph{special local} Kitaev limit, denoted by $H_K$, where $\Delta_{L-1} = w$, $\mu_L = 0$ and $u_{L-1} = 0$ (recall that the $L$'th site refers to the last site of the left wire). In this limit there are interactions throughout all of the wire except the last two sites, which allows a fully decoupled Majorana to sit on the right hand edge, and thus guarantees a precise even-odd degeneracy for all many-body states~\cite{Kells2015}. In the same way, $H_K$ is guaranteed to satisfy the exact TQO condition for operators of size at most $L_K^* = L -1 $, i.e. the entire length of the wire except a single site. The fact that our fully interacting system of interest, $H$, differs from $H_K$ on a single site, means that it is relatively easy to quasi-adiabatically connect the two. This manifests in the approximate TQO condition of $H$~(Eq.~\ref{tqo_assumption} in the main text) as a smaller exponential error than would arise from a longer quasi-adiabatic path to $H$.

Taking $H'=H_K$, we now follow~\cite{Hastings2005} to derive approximate TQO for $H$. Denoting the ground states of $H_K$ as $\ket{a}_K$ (for $a=e,o$), we have the exact TQO condition: $\phantom|_K \hspace{-0.5mm} \bra{e}O\ket{e}_K = \phantom|_K \hspace{-0.5mm} \bra{o}O\ket{o}_K$ for every local operator $O$ supported in a region of size at most $L_K^* = L-1$. We consider a 1-parameter family of Hamiltonians $H(t)$, where $H(0)=H$ and $H(\tau) = H_K$, where $\tau$ is the time we tune for. In~\cite{Hastings2005} they define a unitary operator, denoted here by $V$, that quasi-adiabatically transitions between $H$ and $H_K$. We only consider cases for which $H(t)$ remains gapped throughout the tuning process~\cite{Moon2018}. Since this is a local process, $H(t)$ is always a sum of local terms.

Following~\cite{Hastings2005}, we consider some local observable $O$. We then form $O_l$, which only acts on sites within a distance $l$ from the support of $O$. Specifically, $O_l$ is the partial trace (up to some constant) of $V O V^{\dagger}$ over all sites more than a distance $l$ from the support of $O$. One can then show that,
\begin{equation}
\bra{a}O\ket{a} = \phantom|_K \hspace{-0.5mm} \bra{a} Q^{\dagger} O_l Q \ket{a}_K + \mathcal{O} (e^{-l/\xi} ),
\end{equation}
where $\xi$ is some constant, and $Q$ is a unitary operator acting only within the ground state manifold. 

To use the exact TQO of the local $H_K$ limit in the RHS above, we require $O_l$ to be supported in a region of size at most $L^*_K = L-1$. If the length of $O$ is $d_O$, then the length of $O_l$ is $d_O + 2 l$, which is less than or equal to $L^*_K$ if $l \leq 1/2 ( L- 1 - d_O ) $. To connect up with Eq.~\ref{tqo_assumption} in the main text, we further assume that $d_O \leq L^*$, where $L^*$ is some length satisfying $c L \leq L^* < L$, for some constant $c>0$. At worst we then have $l\sim L$, and hence the error above is order $\mathcal{O} (e^{-L/\xi} )$, for some (possibly different) constant $\xi$.

Finally, given the exact TQO of the special local Kitaev limit, and the fact that $Q$ only acts within the ground state manifold, we find that the expectation value, $\phantom|_K \hspace{-0.5mm} \bra{a} Q^{\dagger} O_l Q \ket{a}_K$, is the same for both sectors. For the fully interacting Majorana system, $H$, this implies that $\bra{a}O\ket{a}$ is the same for both sectors, up to $\mathcal{O} (e^{-L/\xi} )$ for operators $O$ of size at most $L^*$. This is the approximate TQO condition in Eq.~\ref{tqo_assumption} in the main text. 

While the above argument concerned the special local Kitaev limit and a locally connected $H$, it is clearly applicable to any pair of locally connected (gapped) systems where one is known to satisfy an exact TQO condition. This is of course also true more generally for higher dimensional ground state manifolds satisfying the same condition. 

\section{Extension to multi-particle Green's functions}\label{app:MultPartGF}

Here we show some results for the 2-particle spectral densities. We set $e^{i \delta'}  = 2 c^\dagger_{x'}c_{x'} -\mathds{1}$ and $e^{-i \delta}  =  2 c^\dagger_{x} c_{x}-\mathds{1} $ in $U(t)$ in Eq.~\ref{eq:interaction_picture_unitary} in the main text and define the Green's function for the $a$-sector as
\be
G^{a}(x,x',t) := e^{i E_0 t } \langle a | U(t) |a \rangle \; ,
\ee
which depends on the operator positions, $x$ and $x'$, and the time $t$. We also fix $x=L/2$ (the middle of the wire) so as to not couple to the zero-modes (which tends to blur our spatial Fourier transform) and then plot $G^{e}(L/2,x',t) $ and the difference $D:=G^{e}(L/2,x',t)-G^{o}(L/2,x',t)$, which is given in a log scale for clarity. Fig.~\ref{fig:LD} shows, up to numerical errors of  $\mathcal{O}(10^{-8})$ that $D$ is exponentially suppressed for some initial time $T^*$. Furthermore, this time scales with the system size $L$. 

Fig.~\ref{fig:FTG} shows the shape of the 2-particle spectral density as a function of the total momentum, after cutting off the Green's function for $t> T^*$. Fig~\ref{fig:FTG} can be compared against the exact density determined from the single particle solutions of a periodic system, given in Fig.~\ref{fig:TPD}.

The energetic resolution obtained in Fig.~\ref{fig:FTG} is effectively governed by the Lieb-Robinson velocity and $L$.  However, as we stressed above in the constraints section, we are free to choose from a large number of local operators with which we can generate a much larger set of spectral constraints~(Eq. \ref{eq:spectral_diff_bound}~above). For example, we could generate the entire 3-parameter functions 
$G^{a}(x,x',t)$ and perform a 2D Fourier transform on the first two spatial components, from which we could obtain a tighter $k$-space resolution. In further Supplementary Material we provide two animated GIF files showing how this affects things for different momentum slices for the even and odd sectors. While the methodology still suffers from numerical artifacts, our key point is that movies are the same up to exponentially small corrections when the cut-off $T^*$ is enforced by an appropriate choice of $\Theta(t)$.

\bibliography{refs.bib}

%merlin.mbs apsrev4-1.bst 2010-07-25 4.21a (PWD, AO, DPC) hacked
%Control: key (0)
%Control: author (0) dotless jnrlst
%Control: editor formatted (1) identically to author
%Control: production of article title (0) allowed
%Control: page (1) range
%Control: year (0) verbatim
%Control: production of eprint (0) enabled
\begin{thebibliography}{77}%
\makeatletter
\providecommand \@ifxundefined [1]{%
 \@ifx{#1\undefined}
}%
\providecommand \@ifnum [1]{%
 \ifnum #1\expandafter \@firstoftwo
 \else \expandafter \@secondoftwo
 \fi
}%
\providecommand \@ifx [1]{%
 \ifx #1\expandafter \@firstoftwo
 \else \expandafter \@secondoftwo
 \fi
}%
\providecommand \natexlab [1]{#1}%
\providecommand \enquote  [1]{``#1''}%
\providecommand \bibnamefont  [1]{#1}%
\providecommand \bibfnamefont [1]{#1}%
\providecommand \citenamefont [1]{#1}%
\providecommand \href@noop [0]{\@secondoftwo}%
\providecommand \href [0]{\begingroup \@sanitize@url \@href}%
\providecommand \@href[1]{\@@startlink{#1}\@@href}%
\providecommand \@@href[1]{\endgroup#1\@@endlink}%
\providecommand \@sanitize@url [0]{\catcode `\\12\catcode `\$12\catcode
  `\&12\catcode `\#12\catcode `\^12\catcode `\_12\catcode `\%12\relax}%
\providecommand \@@startlink[1]{}%
\providecommand \@@endlink[0]{}%
\providecommand \url  [0]{\begingroup\@sanitize@url \@url }%
\providecommand \@url [1]{\endgroup\@href {#1}{\urlprefix }}%
\providecommand \urlprefix  [0]{URL }%
\providecommand \Eprint [0]{\href }%
\providecommand \doibase [0]{http://dx.doi.org/}%
\providecommand \selectlanguage [0]{\@gobble}%
\providecommand \bibinfo  [0]{\@secondoftwo}%
\providecommand \bibfield  [0]{\@secondoftwo}%
\providecommand \translation [1]{[#1]}%
\providecommand \BibitemOpen [0]{}%
\providecommand \bibitemStop [0]{}%
\providecommand \bibitemNoStop [0]{.\EOS\space}%
\providecommand \EOS [0]{\spacefactor3000\relax}%
\providecommand \BibitemShut  [1]{\csname bibitem#1\endcsname}%
\let\auto@bib@innerbib\@empty
%</preamble>
\bibitem [{\citenamefont {Kitaev}(2001)}]{Kitaev2001}%
  \BibitemOpen
  \bibfield  {author} {\bibinfo {author} {\bibfnamefont {A.~Y.}\ \bibnamefont
  {Kitaev}},\ }\bibfield  {title} {\enquote {\bibinfo {title} {Unpaired
  majorana fermions in quantum wires},}\ }\href
  {http://stacks.iop.org/1063-7869/44/i=10S/a=S29} {\bibfield  {journal}
  {\bibinfo  {journal} {Physics-Uspekhi}\ }\textbf {\bibinfo {volume} {44}},\
  \bibinfo {pages} {131} (\bibinfo {year} {2001})}\BibitemShut {NoStop}%
\bibitem [{\citenamefont {Dennis}\ \emph {et~al.}(2002)\citenamefont {Dennis},
  \citenamefont {Kitaev}, \citenamefont {Landahl},\ and\ \citenamefont
  {Preskill}}]{Dennis2002}%
  \BibitemOpen
  \bibfield  {author} {\bibinfo {author} {\bibfnamefont {E.}~\bibnamefont
  {Dennis}}, \bibinfo {author} {\bibfnamefont {A.~Y.}\ \bibnamefont {Kitaev}},
  \bibinfo {author} {\bibfnamefont {A.}~\bibnamefont {Landahl}}, \ and\
  \bibinfo {author} {\bibfnamefont {J.}~\bibnamefont {Preskill}},\ }\bibfield
  {title} {\enquote {\bibinfo {title} {Topological quantum memory},}\ }\href
  {\doibase 10.1063/1.1499754} {\bibfield  {journal} {\bibinfo  {journal}
  {Journal of Mathematical Physics}\ }\textbf {\bibinfo {volume} {43}},\
  \bibinfo {pages} {4452--4505} (\bibinfo {year} {2002})},\ \Eprint
  {http://arxiv.org/abs/https://doi.org/10.1063/1.1499754}
  {https://doi.org/10.1063/1.1499754} \BibitemShut {NoStop}%
\bibitem [{\citenamefont {Kitaev}(2003)}]{Kitaev2003}%
  \BibitemOpen
  \bibfield  {author} {\bibinfo {author} {\bibfnamefont {A.~Y.}\ \bibnamefont
  {Kitaev}},\ }\bibfield  {title} {\enquote {\bibinfo {title} {Fault-tolerant
  quantum computation by anyons},}\ }\href {\doibase
  https://doi.org/10.1016/S0003-4916(02)00018-0} {\bibfield  {journal}
  {\bibinfo  {journal} {Annals of Physics}\ }\textbf {\bibinfo {volume}
  {303}},\ \bibinfo {pages} {2 -- 30} (\bibinfo {year} {2003})}\BibitemShut
  {NoStop}%
\bibitem [{\citenamefont {Kitaev}(2006)}]{Kitaev2006}%
  \BibitemOpen
  \bibfield  {author} {\bibinfo {author} {\bibfnamefont {A.~Y.}\ \bibnamefont
  {Kitaev}},\ }\bibfield  {title} {\enquote {\bibinfo {title} {Anyons in an
  exactly solved model and beyond},}\ }\href {\doibase
  https://doi.org/10.1016/j.aop.2005.10.005} {\bibfield  {journal} {\bibinfo
  {journal} {Annals of Physics}\ }\textbf {\bibinfo {volume} {321}},\ \bibinfo
  {pages} {2 -- 111} (\bibinfo {year} {2006})}\BibitemShut {NoStop}%
\bibitem [{\citenamefont {Nayak}\ \emph {et~al.}(2008)\citenamefont {Nayak},
  \citenamefont {Simon}, \citenamefont {Stern}, \citenamefont {Freedman},\ and\
  \citenamefont {Das~Sarma}}]{NayakRev}%
  \BibitemOpen
  \bibfield  {author} {\bibinfo {author} {\bibfnamefont {C.}~\bibnamefont
  {Nayak}}, \bibinfo {author} {\bibfnamefont {S.~H.}\ \bibnamefont {Simon}},
  \bibinfo {author} {\bibfnamefont {A.}~\bibnamefont {Stern}}, \bibinfo
  {author} {\bibfnamefont {M.}~\bibnamefont {Freedman}}, \ and\ \bibinfo
  {author} {\bibfnamefont {S.}~\bibnamefont {Das~Sarma}},\ }\bibfield  {title}
  {\enquote {\bibinfo {title} {Non-abelian anyons and topological quantum
  computation},}\ }\href {\doibase 10.1103/RevModPhys.80.1083} {\bibfield
  {journal} {\bibinfo  {journal} {Rev. Mod. Phys.}\ }\textbf {\bibinfo {volume}
  {80}},\ \bibinfo {pages} {1083--1159} (\bibinfo {year} {2008})}\BibitemShut
  {NoStop}%
\bibitem [{\citenamefont {Stanescu}\ and\ \citenamefont
  {Tewari}(2013)}]{Stanescu2013}%
  \BibitemOpen
  \bibfield  {author} {\bibinfo {author} {\bibfnamefont {T.~D.}\ \bibnamefont
  {Stanescu}}\ and\ \bibinfo {author} {\bibfnamefont {S.}~\bibnamefont
  {Tewari}},\ }\bibfield  {title} {\enquote {\bibinfo {title} {Majorana
  fermions in semiconductor nanowires: fundamentals, modeling, and
  experiment},}\ }\href {\doibase 10.1088/0953-8984/25/23/233201} {\bibfield
  {journal} {\bibinfo  {journal} {Journal of Physics: Condensed Matter}\
  }\textbf {\bibinfo {volume} {25}},\ \bibinfo {pages} {233201} (\bibinfo
  {year} {2013})}\BibitemShut {NoStop}%
\bibitem [{\citenamefont {Wen}\ and\ \citenamefont
  {Niu}(1990)}]{PhysRevB.41.9377}%
  \BibitemOpen
  \bibfield  {author} {\bibinfo {author} {\bibfnamefont {X.~G.}\ \bibnamefont
  {Wen}}\ and\ \bibinfo {author} {\bibfnamefont {Q.}~\bibnamefont {Niu}},\
  }\bibfield  {title} {\enquote {\bibinfo {title} {Ground-state degeneracy of
  the fractional quantum hall states in the presence of a random potential and
  on high-genus riemann surfaces},}\ }\href {\doibase 10.1103/PhysRevB.41.9377}
  {\bibfield  {journal} {\bibinfo  {journal} {Phys. Rev. B}\ }\textbf {\bibinfo
  {volume} {41}},\ \bibinfo {pages} {9377--9396} (\bibinfo {year}
  {1990})}\BibitemShut {NoStop}%
\bibitem [{\citenamefont {Wen}(1990)}]{Wen:1989iv}%
  \BibitemOpen
  \bibfield  {author} {\bibinfo {author} {\bibfnamefont {X.G.}\ \bibnamefont
  {Wen}},\ }\bibfield  {title} {\enquote {\bibinfo {title} {{Topological Order
  in Rigid States}},}\ }\href {\doibase 10.1142/S0217979290000139} {\bibfield
  {journal} {\bibinfo  {journal} {Int. J. Mod. Phys. B}\ }\textbf {\bibinfo
  {volume} {4}},\ \bibinfo {pages} {239} (\bibinfo {year} {1990})}\BibitemShut
  {NoStop}%
\bibitem [{\citenamefont {Hastings}\ and\ \citenamefont
  {Wen}(2005)}]{Hastings2005}%
  \BibitemOpen
  \bibfield  {author} {\bibinfo {author} {\bibfnamefont {M.~B.}\ \bibnamefont
  {Hastings}}\ and\ \bibinfo {author} {\bibfnamefont {X.~G.}\ \bibnamefont
  {Wen}},\ }\bibfield  {title} {\enquote {\bibinfo {title} {Quasiadiabatic
  continuation of quantum states: The stability of topological ground-state
  degeneracy and emergent gauge invariance},}\ }\href {\doibase
  10.1103/PhysRevB.72.045141} {\bibfield  {journal} {\bibinfo  {journal} {Phys.
  Rev. B}\ }\textbf {\bibinfo {volume} {72}},\ \bibinfo {pages} {045141}
  (\bibinfo {year} {2005})}\BibitemShut {NoStop}%
\bibitem [{\citenamefont {Bravyi}\ \emph {et~al.}(2010)\citenamefont {Bravyi},
  \citenamefont {Hastings},\ and\ \citenamefont {Michalakis}}]{Bravyi2010h}%
  \BibitemOpen
  \bibfield  {author} {\bibinfo {author} {\bibfnamefont {S.}~\bibnamefont
  {Bravyi}}, \bibinfo {author} {\bibfnamefont {M.~B.}\ \bibnamefont
  {Hastings}}, \ and\ \bibinfo {author} {\bibfnamefont {S.}~\bibnamefont
  {Michalakis}},\ }\bibfield  {title} {\enquote {\bibinfo {title} {Topological
  quantum order: Stability under local perturbations},}\ }\href {\doibase
  10.1063/1.3490195} {\bibfield  {journal} {\bibinfo  {journal} {Journal of
  Mathematical Physics}\ }\textbf {\bibinfo {volume} {51}},\ \bibinfo {pages}
  {093512} (\bibinfo {year} {2010})},\ \Eprint
  {http://arxiv.org/abs/https://doi.org/10.1063/1.3490195}
  {https://doi.org/10.1063/1.3490195} \BibitemShut {NoStop}%
\bibitem [{\citenamefont {Zeng}\ \emph {et~al.}(2019)\citenamefont {Zeng},
  \citenamefont {Chen}, \citenamefont {Zhou},\ and\ \citenamefont
  {Wen}}]{zeng2019quantum}%
  \BibitemOpen
  \bibfield  {author} {\bibinfo {author} {\bibfnamefont {B.}~\bibnamefont
  {Zeng}}, \bibinfo {author} {\bibfnamefont {X.}~\bibnamefont {Chen}}, \bibinfo
  {author} {\bibfnamefont {D.L.}\ \bibnamefont {Zhou}}, \ and\ \bibinfo
  {author} {\bibfnamefont {X.G.}\ \bibnamefont {Wen}},\ }\href
  {https://books.google.nl/books?id=kmqPDwAAQBAJ} {\emph {\bibinfo {title}
  {Quantum Information Meets Quantum Matter: From Quantum Entanglement to
  Topological Phases of Many-Body Systems}}},\ Quantum Science and Technology\
  (\bibinfo  {publisher} {Springer New York},\ \bibinfo {year}
  {2019})\BibitemShut {NoStop}%
\bibitem [{\citenamefont {Conlon}\ \emph {et~al.}(2019)\citenamefont {Conlon},
  \citenamefont {Pellegrino}, \citenamefont {Slingerland}, \citenamefont
  {Dooley},\ and\ \citenamefont {Kells}}]{Conlon2019}%
  \BibitemOpen
  \bibfield  {author} {\bibinfo {author} {\bibfnamefont {A.}~\bibnamefont
  {Conlon}}, \bibinfo {author} {\bibfnamefont {D.}~\bibnamefont {Pellegrino}},
  \bibinfo {author} {\bibfnamefont {J.~K.}\ \bibnamefont {Slingerland}},
  \bibinfo {author} {\bibfnamefont {S.}~\bibnamefont {Dooley}}, \ and\ \bibinfo
  {author} {\bibfnamefont {G.}~\bibnamefont {Kells}},\ }\bibfield  {title}
  {\enquote {\bibinfo {title} {Error generation and propagation in
  majorana-based topological qubits},}\ }\href {\doibase
  10.1103/PhysRevB.100.134307} {\bibfield  {journal} {\bibinfo  {journal}
  {Phys. Rev. B}\ }\textbf {\bibinfo {volume} {100}},\ \bibinfo {pages}
  {134307} (\bibinfo {year} {2019})}\BibitemShut {NoStop}%
\bibitem [{\citenamefont {Fu}\ and\ \citenamefont {Kane}(2008)}]{Fu2008}%
  \BibitemOpen
  \bibfield  {author} {\bibinfo {author} {\bibfnamefont {L.}~\bibnamefont
  {Fu}}\ and\ \bibinfo {author} {\bibfnamefont {C.~L.}\ \bibnamefont {Kane}},\
  }\bibfield  {title} {\enquote {\bibinfo {title} {Superconducting proximity
  effect and majorana fermions at the surface of a topological insulator},}\
  }\href {\doibase 10.1103/PhysRevLett.100.096407} {\bibfield  {journal}
  {\bibinfo  {journal} {Phys. Rev. Lett.}\ }\textbf {\bibinfo {volume} {100}},\
  \bibinfo {pages} {096407} (\bibinfo {year} {2008})}\BibitemShut {NoStop}%
\bibitem [{\citenamefont {Lutchyn}\ \emph {et~al.}(2010)\citenamefont
  {Lutchyn}, \citenamefont {Sau},\ and\ \citenamefont
  {Das~Sarma}}]{Lutchyn2010}%
  \BibitemOpen
  \bibfield  {author} {\bibinfo {author} {\bibfnamefont {R.~M.}\ \bibnamefont
  {Lutchyn}}, \bibinfo {author} {\bibfnamefont {J.~D.}\ \bibnamefont {Sau}}, \
  and\ \bibinfo {author} {\bibfnamefont {S.}~\bibnamefont {Das~Sarma}},\
  }\bibfield  {title} {\enquote {\bibinfo {title} {Majorana fermions and a
  topological phase transition in semiconductor-superconductor
  heterostructures},}\ }\href {\doibase 10.1103/PhysRevLett.105.077001}
  {\bibfield  {journal} {\bibinfo  {journal} {Phys. Rev. Lett.}\ }\textbf
  {\bibinfo {volume} {105}},\ \bibinfo {pages} {077001} (\bibinfo {year}
  {2010})}\BibitemShut {NoStop}%
\bibitem [{\citenamefont {Oreg}\ \emph {et~al.}(2010)\citenamefont {Oreg},
  \citenamefont {Refael},\ and\ \citenamefont {von Oppen}}]{Oreg2010}%
  \BibitemOpen
  \bibfield  {author} {\bibinfo {author} {\bibfnamefont {Y.}~\bibnamefont
  {Oreg}}, \bibinfo {author} {\bibfnamefont {G.}~\bibnamefont {Refael}}, \ and\
  \bibinfo {author} {\bibfnamefont {F.}~\bibnamefont {von Oppen}},\ }\bibfield
  {title} {\enquote {\bibinfo {title} {Helical liquids and majorana bound
  states in quantum wires},}\ }\href {\doibase 10.1103/PhysRevLett.105.177002}
  {\bibfield  {journal} {\bibinfo  {journal} {Phys. Rev. Lett.}\ }\textbf
  {\bibinfo {volume} {105}},\ \bibinfo {pages} {177002} (\bibinfo {year}
  {2010})}\BibitemShut {NoStop}%
\bibitem [{\citenamefont {Clarke}\ \emph {et~al.}(2011)\citenamefont {Clarke},
  \citenamefont {Sau},\ and\ \citenamefont {Tewari}}]{Sau2011}%
  \BibitemOpen
  \bibfield  {author} {\bibinfo {author} {\bibfnamefont {D.~J.}\ \bibnamefont
  {Clarke}}, \bibinfo {author} {\bibfnamefont {J.~D.}\ \bibnamefont {Sau}}, \
  and\ \bibinfo {author} {\bibfnamefont {S.}~\bibnamefont {Tewari}},\
  }\bibfield  {title} {\enquote {\bibinfo {title} {Majorana fermion exchange in
  quasi-one-dimensional networks},}\ }\href {\doibase
  10.1103/PhysRevB.84.035120} {\bibfield  {journal} {\bibinfo  {journal} {Phys.
  Rev. B}\ }\textbf {\bibinfo {volume} {84}},\ \bibinfo {pages} {035120}
  (\bibinfo {year} {2011})}\BibitemShut {NoStop}%
\bibitem [{\citenamefont {Alicea}\ \emph {et~al.}(2011)\citenamefont {Alicea},
  \citenamefont {Oreg}, \citenamefont {Refael}, \citenamefont {von Oppen},\
  and\ \citenamefont {Fisher}}]{AliceaNat2011}%
  \BibitemOpen
  \bibfield  {author} {\bibinfo {author} {\bibfnamefont {J.}~\bibnamefont
  {Alicea}}, \bibinfo {author} {\bibfnamefont {Y.}~\bibnamefont {Oreg}},
  \bibinfo {author} {\bibfnamefont {G.}~\bibnamefont {Refael}}, \bibinfo
  {author} {\bibfnamefont {F.}~\bibnamefont {von Oppen}}, \ and\ \bibinfo
  {author} {\bibfnamefont {M.~P.~A.}\ \bibnamefont {Fisher}},\ }\bibfield
  {title} {\enquote {\bibinfo {title} {Non-abelian statistics and topological
  quantum information processing in 1d wire networks},}\ }\href@noop {}
  {\bibfield  {journal} {\bibinfo  {journal} {Nature Physics}\ }\textbf
  {\bibinfo {volume} {7}},\ \bibinfo {pages} {412 EP --} (\bibinfo {year}
  {2011})}\BibitemShut {NoStop}%
\bibitem [{\citenamefont {Haegeman}\ \emph {et~al.}(2011)\citenamefont
  {Haegeman}, \citenamefont {Cirac}, \citenamefont {Osborne}, \citenamefont
  {Pi\ifmmode~\check{z}\else \v{z}\fi{}orn}, \citenamefont {Verschelde},\ and\
  \citenamefont {Verstraete}}]{HaegemanTDVP}%
  \BibitemOpen
  \bibfield  {author} {\bibinfo {author} {\bibfnamefont {J.}~\bibnamefont
  {Haegeman}}, \bibinfo {author} {\bibfnamefont {J.~I.}\ \bibnamefont {Cirac}},
  \bibinfo {author} {\bibfnamefont {T.~J.}\ \bibnamefont {Osborne}}, \bibinfo
  {author} {\bibfnamefont {I.}~\bibnamefont {Pi\ifmmode~\check{z}\else
  \v{z}\fi{}orn}}, \bibinfo {author} {\bibfnamefont {H.}~\bibnamefont
  {Verschelde}}, \ and\ \bibinfo {author} {\bibfnamefont {F.}~\bibnamefont
  {Verstraete}},\ }\bibfield  {title} {\enquote {\bibinfo {title}
  {Time-dependent variational principle for quantum lattices},}\ }\href
  {\doibase 10.1103/PhysRevLett.107.070601} {\bibfield  {journal} {\bibinfo
  {journal} {Phys. Rev. Lett.}\ }\textbf {\bibinfo {volume} {107}},\ \bibinfo
  {pages} {070601} (\bibinfo {year} {2011})}\BibitemShut {NoStop}%
\bibitem [{\citenamefont {Paeckel}\ \emph {et~al.}(2019)\citenamefont
  {Paeckel}, \citenamefont {K{\"o}hler}, \citenamefont {Swoboda}, \citenamefont
  {Manmana}, \citenamefont {Schollw{\"o}ck},\ and\ \citenamefont
  {Hubig}}]{Paeckel2019}%
  \BibitemOpen
  \bibfield  {author} {\bibinfo {author} {\bibfnamefont {S.}~\bibnamefont
  {Paeckel}}, \bibinfo {author} {\bibfnamefont {T.}~\bibnamefont {K{\"o}hler}},
  \bibinfo {author} {\bibfnamefont {A.}~\bibnamefont {Swoboda}}, \bibinfo
  {author} {\bibfnamefont {S.R.}\ \bibnamefont {Manmana}}, \bibinfo {author}
  {\bibfnamefont {U.}~\bibnamefont {Schollw{\"o}ck}}, \ and\ \bibinfo {author}
  {\bibfnamefont {C.}~\bibnamefont {Hubig}},\ }\bibfield  {title} {\enquote
  {\bibinfo {title} {Time-evolution methods for matrix-product states},}\
  }\href {\doibase https://doi.org/10.1016/j.aop.2019.167998} {\bibfield
  {journal} {\bibinfo  {journal} {Annals of Physics}\ }\textbf {\bibinfo
  {volume} {411}},\ \bibinfo {pages} {167998} (\bibinfo {year}
  {2019})}\BibitemShut {NoStop}%
\bibitem [{Note1()}]{Note1}%
  \BibitemOpen
  \bibinfo {note} {By \protect \emph {local Hamiltonian} we mean a sum of
  Hermitian operators that are each only non-trivial on several neighbouring
  sites.}\BibitemShut {Stop}%
\bibitem [{Note2()}]{Note2}%
  \BibitemOpen
  \bibinfo {note} {This is the case for the Majorana example below. It is
  straightforward to generalise the argument to a higher dimensional ground
  state manifold.}\BibitemShut {Stop}%
\bibitem [{\citenamefont {Bravyi}\ and\ \citenamefont
  {Hastings}(2011)}]{Bravyi2011h}%
  \BibitemOpen
  \bibfield  {author} {\bibinfo {author} {\bibfnamefont {S.}~\bibnamefont
  {Bravyi}}\ and\ \bibinfo {author} {\bibfnamefont {M.~B.}\ \bibnamefont
  {Hastings}},\ }\bibfield  {title} {\enquote {\bibinfo {title} {A short proof
  of stability of topological order under local perturbations},}\ }\href
  {\doibase 10.1007/s00220-011-1346-2} {\bibfield  {journal} {\bibinfo
  {journal} {Communications in Mathematical Physics}\ }\textbf {\bibinfo
  {volume} {307}},\ \bibinfo {pages} {609} (\bibinfo {year}
  {2011})}\BibitemShut {NoStop}%
\bibitem [{Note3()}]{Note3}%
  \BibitemOpen
  \bibinfo {note} {$H$ is a sum of a polynomial (in $L$) number of $k$-local
  terms, $H_{x}$. TQO implies that {$ \langle a|H_{x} |a \rangle $} is sector
  independent up to $\protect \mathcal {O}(e^{-L/\xi })$, which dominates the
  polynomial in $L$ in the sum {$\langle a| H |a \rangle = \DOTSB \sum@
  \slimits@ _x \langle a|H_{x}|a \rangle $}.}\BibitemShut {Stop}%
\bibitem [{\citenamefont {Lieb}\ and\ \citenamefont
  {Robinson}(1972)}]{Lieb:1972wy}%
  \BibitemOpen
  \bibfield  {author} {\bibinfo {author} {\bibfnamefont {E.H.}\ \bibnamefont
  {Lieb}}\ and\ \bibinfo {author} {\bibfnamefont {D.W.}\ \bibnamefont
  {Robinson}},\ }\bibfield  {title} {\enquote {\bibinfo {title} {{The finite
  group velocity of quantum spin systems}},}\ }\href {\doibase
  10.1007/BF01645779} {\bibfield  {journal} {\bibinfo  {journal} {Commun. Math.
  Phys.}\ }\textbf {\bibinfo {volume} {28}},\ \bibinfo {pages} {251--257}
  (\bibinfo {year} {1972})}\BibitemShut {NoStop}%
\bibitem [{Note4()}]{Note4}%
  \BibitemOpen
  \bibinfo {note} {Our argument involves several exponential error terms; each
  with its own rate constant $\xi $. To avoid ambiguity, we take $\xi $ to be
  the smallest such constant such that each error estimate is
  valid.}\BibitemShut {Stop}%
\bibitem [{\citenamefont {Hastings}(2004)}]{Hastings_2004}%
  \BibitemOpen
  \bibfield  {author} {\bibinfo {author} {\bibfnamefont {M.~B.}\ \bibnamefont
  {Hastings}},\ }\bibfield  {title} {\enquote {\bibinfo {title} {Locality in
  quantum and markov dynamics on lattices and networks},}\ }\href {\doibase
  10.1103/physrevlett.93.140402} {\bibfield  {journal} {\bibinfo  {journal}
  {Physical Review Letters}\ }\textbf {\bibinfo {volume} {93}} (\bibinfo {year}
  {2004}),\ 10.1103/physrevlett.93.140402}\BibitemShut {NoStop}%
\bibitem [{\citenamefont {Hastings}\ and\ \citenamefont
  {Koma}(2006)}]{Hastings:2005pr}%
  \BibitemOpen
  \bibfield  {author} {\bibinfo {author} {\bibfnamefont {M.~B.}\ \bibnamefont
  {Hastings}}\ and\ \bibinfo {author} {\bibfnamefont {T.}~\bibnamefont
  {Koma}},\ }\bibfield  {title} {\enquote {\bibinfo {title} {{Spectral gap and
  exponential decay of correlations}},}\ }\href {\doibase
  10.1007/s00220-006-0030-4} {\bibfield  {journal} {\bibinfo  {journal}
  {Commun. Math. Phys.}\ }\textbf {\bibinfo {volume} {265}},\ \bibinfo {pages}
  {781--804} (\bibinfo {year} {2006})},\ \Eprint
  {http://arxiv.org/abs/math-ph/0507008} {arXiv:math-ph/0507008} \BibitemShut
  {NoStop}%
\bibitem [{\citenamefont {Mourik}\ \emph {et~al.}(2012)\citenamefont {Mourik},
  \citenamefont {Zuo}, \citenamefont {Frolov}, \citenamefont {Plissard},
  \citenamefont {Bakkers},\ and\ \citenamefont
  {Kouwenhoven}}]{Kouwenhoven2012}%
  \BibitemOpen
  \bibfield  {author} {\bibinfo {author} {\bibfnamefont {V.}~\bibnamefont
  {Mourik}}, \bibinfo {author} {\bibfnamefont {K.}~\bibnamefont {Zuo}},
  \bibinfo {author} {\bibfnamefont {S.~M.}\ \bibnamefont {Frolov}}, \bibinfo
  {author} {\bibfnamefont {S.~R.}\ \bibnamefont {Plissard}}, \bibinfo {author}
  {\bibfnamefont {E.~P. A.~M.}\ \bibnamefont {Bakkers}}, \ and\ \bibinfo
  {author} {\bibfnamefont {L.~P.}\ \bibnamefont {Kouwenhoven}},\ }\bibfield
  {title} {\enquote {\bibinfo {title} {Signatures of majorana fermions in
  hybrid superconductor-semiconductor nanowire devices},}\ }\href {\doibase
  10.1126/science.1222360} {\bibfield  {journal} {\bibinfo  {journal}
  {Science}\ }\textbf {\bibinfo {volume} {336}},\ \bibinfo {pages} {1003--1007}
  (\bibinfo {year} {2012})}\BibitemShut {NoStop}%
\bibitem [{\citenamefont {Deng}\ \emph {et~al.}(2012)\citenamefont {Deng},
  \citenamefont {Yu}, \citenamefont {Huang}, \citenamefont {Larsson},
  \citenamefont {Caroff},\ and\ \citenamefont {Xu}}]{Deng2012}%
  \BibitemOpen
  \bibfield  {author} {\bibinfo {author} {\bibfnamefont {M.~T.}\ \bibnamefont
  {Deng}}, \bibinfo {author} {\bibfnamefont {C.~L.}\ \bibnamefont {Yu}},
  \bibinfo {author} {\bibfnamefont {G.~Y.}\ \bibnamefont {Huang}}, \bibinfo
  {author} {\bibfnamefont {M.}~\bibnamefont {Larsson}}, \bibinfo {author}
  {\bibfnamefont {P.}~\bibnamefont {Caroff}}, \ and\ \bibinfo {author}
  {\bibfnamefont {H.~Q.}\ \bibnamefont {Xu}},\ }\bibfield  {title} {\enquote
  {\bibinfo {title} {Anomalous zero-bias conductance peak in a nb--insb
  nanowire--nb hybrid device},}\ }\bibfield  {booktitle} {\emph {\bibinfo
  {booktitle} {Nano Letters}},\ }\href {\doibase 10.1021/nl303758w} {\bibfield
  {journal} {\bibinfo  {journal} {Nano Letters}\ }\textbf {\bibinfo {volume}
  {12}},\ \bibinfo {pages} {6414--6419} (\bibinfo {year} {2012})}\BibitemShut
  {NoStop}%
\bibitem [{\citenamefont {Das}\ \emph {et~al.}(2012)\citenamefont {Das},
  \citenamefont {Ronen}, \citenamefont {Most}, \citenamefont {Oreg},
  \citenamefont {Heiblum},\ and\ \citenamefont {Shtrikman}}]{Das2012}%
  \BibitemOpen
  \bibfield  {author} {\bibinfo {author} {\bibfnamefont {A.}~\bibnamefont
  {Das}}, \bibinfo {author} {\bibfnamefont {Y.}~\bibnamefont {Ronen}}, \bibinfo
  {author} {\bibfnamefont {Y.}~\bibnamefont {Most}}, \bibinfo {author}
  {\bibfnamefont {Y.}~\bibnamefont {Oreg}}, \bibinfo {author} {\bibfnamefont
  {M.}~\bibnamefont {Heiblum}}, \ and\ \bibinfo {author} {\bibfnamefont
  {H.}~\bibnamefont {Shtrikman}},\ }\bibfield  {title} {\enquote {\bibinfo
  {title} {Zero-bias peaks and splitting in an al--inas nanowire topological
  superconductor as a signature of majorana fermions},}\ }\href
  {https://doi.org/10.1038/nphys2479} {\bibfield  {journal} {\bibinfo
  {journal} {Nature Physics}\ }\textbf {\bibinfo {volume} {8}},\ \bibinfo
  {pages} {887 EP --} (\bibinfo {year} {2012})}\BibitemShut {NoStop}%
\bibitem [{\citenamefont {Finck}\ \emph {et~al.}(2013)\citenamefont {Finck},
  \citenamefont {Van~Harlingen}, \citenamefont {Mohseni}, \citenamefont
  {Jung},\ and\ \citenamefont {Li}}]{Finck2013}%
  \BibitemOpen
  \bibfield  {author} {\bibinfo {author} {\bibfnamefont {A.~D.~K.}\
  \bibnamefont {Finck}}, \bibinfo {author} {\bibfnamefont {D.~J.}\ \bibnamefont
  {Van~Harlingen}}, \bibinfo {author} {\bibfnamefont {P.~K.}\ \bibnamefont
  {Mohseni}}, \bibinfo {author} {\bibfnamefont {K.}~\bibnamefont {Jung}}, \
  and\ \bibinfo {author} {\bibfnamefont {X.}~\bibnamefont {Li}},\ }\bibfield
  {title} {\enquote {\bibinfo {title} {Anomalous modulation of a zero-bias peak
  in a hybrid nanowire-superconductor device},}\ }\href {\doibase
  10.1103/PhysRevLett.110.126406} {\bibfield  {journal} {\bibinfo  {journal}
  {Phys. Rev. Lett.}\ }\textbf {\bibinfo {volume} {110}},\ \bibinfo {pages}
  {126406} (\bibinfo {year} {2013})}\BibitemShut {NoStop}%
\bibitem [{\citenamefont {Churchill}\ \emph {et~al.}(2013)\citenamefont
  {Churchill}, \citenamefont {Fatemi}, \citenamefont {Grove-Rasmussen},
  \citenamefont {Deng}, \citenamefont {Caroff}, \citenamefont {Xu},\ and\
  \citenamefont {Marcus}}]{Churchill2013}%
  \BibitemOpen
  \bibfield  {author} {\bibinfo {author} {\bibfnamefont {H.~O.~H.}\
  \bibnamefont {Churchill}}, \bibinfo {author} {\bibfnamefont {V.}~\bibnamefont
  {Fatemi}}, \bibinfo {author} {\bibfnamefont {K.}~\bibnamefont
  {Grove-Rasmussen}}, \bibinfo {author} {\bibfnamefont {M.~T.}\ \bibnamefont
  {Deng}}, \bibinfo {author} {\bibfnamefont {P.}~\bibnamefont {Caroff}},
  \bibinfo {author} {\bibfnamefont {H.~Q.}\ \bibnamefont {Xu}}, \ and\ \bibinfo
  {author} {\bibfnamefont {C.~M.}\ \bibnamefont {Marcus}},\ }\bibfield  {title}
  {\enquote {\bibinfo {title} {Superconductor-nanowire devices from tunneling
  to the multichannel regime: Zero-bias oscillations and magnetoconductance
  crossover},}\ }\href {\doibase 10.1103/PhysRevB.87.241401} {\bibfield
  {journal} {\bibinfo  {journal} {Phys. Rev. B}\ }\textbf {\bibinfo {volume}
  {87}},\ \bibinfo {pages} {241401} (\bibinfo {year} {2013})}\BibitemShut
  {NoStop}%
\bibitem [{\citenamefont {Albrecht}\ \emph {et~al.}(2016)\citenamefont
  {Albrecht}, \citenamefont {Higginbotham}, \citenamefont {Madsen},
  \citenamefont {Kuemmeth}, \citenamefont {Jespersen}, \citenamefont
  {Nyg{\aa}rd}, \citenamefont {Krogstrup},\ and\ \citenamefont
  {Marcus}}]{Albrecht2016}%
  \BibitemOpen
  \bibfield  {author} {\bibinfo {author} {\bibfnamefont {S.~M.}\ \bibnamefont
  {Albrecht}}, \bibinfo {author} {\bibfnamefont {A.~P.}\ \bibnamefont
  {Higginbotham}}, \bibinfo {author} {\bibfnamefont {M.}~\bibnamefont
  {Madsen}}, \bibinfo {author} {\bibfnamefont {F.}~\bibnamefont {Kuemmeth}},
  \bibinfo {author} {\bibfnamefont {T.~S.}\ \bibnamefont {Jespersen}}, \bibinfo
  {author} {\bibfnamefont {J.}~\bibnamefont {Nyg{\aa}rd}}, \bibinfo {author}
  {\bibfnamefont {P.}~\bibnamefont {Krogstrup}}, \ and\ \bibinfo {author}
  {\bibfnamefont {C.~M.}\ \bibnamefont {Marcus}},\ }\bibfield  {title}
  {\enquote {\bibinfo {title} {Exponential protection of zero modes in majorana
  islands},}\ }\href {https://doi.org/10.1038/nature17162} {\bibfield
  {journal} {\bibinfo  {journal} {Nature}\ }\textbf {\bibinfo {volume} {531}},\
  \bibinfo {pages} {206 EP --} (\bibinfo {year} {2016})}\BibitemShut {NoStop}%
\bibitem [{\citenamefont {He}\ \emph {et~al.}(2017)\citenamefont {He},
  \citenamefont {Pan}, \citenamefont {Stern}, \citenamefont {Burks},
  \citenamefont {Che}, \citenamefont {Yin}, \citenamefont {Wang}, \citenamefont
  {Lian}, \citenamefont {Zhou}, \citenamefont {Choi}, \citenamefont {Murata},
  \citenamefont {Kou}, \citenamefont {Chen}, \citenamefont {Nie}, \citenamefont
  {Shao}, \citenamefont {Fan}, \citenamefont {Zhang}, \citenamefont {Liu},
  \citenamefont {Xia},\ and\ \citenamefont {Wang}}]{Zhang2016}%
  \BibitemOpen
  \bibfield  {author} {\bibinfo {author} {\bibfnamefont {Q.~Lin}\ \bibnamefont
  {He}}, \bibinfo {author} {\bibfnamefont {L.}~\bibnamefont {Pan}}, \bibinfo
  {author} {\bibfnamefont {A.~L.}\ \bibnamefont {Stern}}, \bibinfo {author}
  {\bibfnamefont {E.~C.}\ \bibnamefont {Burks}}, \bibinfo {author}
  {\bibfnamefont {X.}~\bibnamefont {Che}}, \bibinfo {author} {\bibfnamefont
  {G.}~\bibnamefont {Yin}}, \bibinfo {author} {\bibfnamefont {J.}~\bibnamefont
  {Wang}}, \bibinfo {author} {\bibfnamefont {B.}~\bibnamefont {Lian}}, \bibinfo
  {author} {\bibfnamefont {Q.}~\bibnamefont {Zhou}}, \bibinfo {author}
  {\bibfnamefont {E.~S.}\ \bibnamefont {Choi}}, \bibinfo {author}
  {\bibfnamefont {K.}~\bibnamefont {Murata}}, \bibinfo {author} {\bibfnamefont
  {X.}~\bibnamefont {Kou}}, \bibinfo {author} {\bibfnamefont {Z.}~\bibnamefont
  {Chen}}, \bibinfo {author} {\bibfnamefont {T.}~\bibnamefont {Nie}}, \bibinfo
  {author} {\bibfnamefont {Q.}~\bibnamefont {Shao}}, \bibinfo {author}
  {\bibfnamefont {Y.}~\bibnamefont {Fan}}, \bibinfo {author} {\bibfnamefont
  {S.~C.}\ \bibnamefont {Zhang}}, \bibinfo {author} {\bibfnamefont
  {F.}~\bibnamefont {Liu}}, \bibinfo {author} {\bibfnamefont {J.}~\bibnamefont
  {Xia}}, \ and\ \bibinfo {author} {\bibfnamefont {K.~L.}\ \bibnamefont
  {Wang}},\ }\bibfield  {title} {\enquote {\bibinfo {title} {Chiral majorana
  fermion modes in a quantum anomalous hall
  insulator{\textendash}superconductor structure},}\ }\href {\doibase
  10.1126/science.aag2792} {\bibfield  {journal} {\bibinfo  {journal}
  {Science}\ }\textbf {\bibinfo {volume} {357}},\ \bibinfo {pages} {294--299}
  (\bibinfo {year} {2017})}\BibitemShut {NoStop}%
\bibitem [{\citenamefont {Deng}\ \emph {et~al.}(2016)\citenamefont {Deng},
  \citenamefont {Vaitiekenas}, \citenamefont {Hansen}, \citenamefont {Danon},
  \citenamefont {Leijnse}, \citenamefont {Flensberg}, \citenamefont
  {Nyg{\aa}rd}, \citenamefont {Krogstrup},\ and\ \citenamefont
  {Marcus}}]{Deng2016}%
  \BibitemOpen
  \bibfield  {author} {\bibinfo {author} {\bibfnamefont {M.~T.}\ \bibnamefont
  {Deng}}, \bibinfo {author} {\bibfnamefont {S.}~\bibnamefont {Vaitiekenas}},
  \bibinfo {author} {\bibfnamefont {E.~B.}\ \bibnamefont {Hansen}}, \bibinfo
  {author} {\bibfnamefont {J.}~\bibnamefont {Danon}}, \bibinfo {author}
  {\bibfnamefont {M.}~\bibnamefont {Leijnse}}, \bibinfo {author} {\bibfnamefont
  {K.}~\bibnamefont {Flensberg}}, \bibinfo {author} {\bibfnamefont
  {J.}~\bibnamefont {Nyg{\aa}rd}}, \bibinfo {author} {\bibfnamefont
  {P.}~\bibnamefont {Krogstrup}}, \ and\ \bibinfo {author} {\bibfnamefont
  {C.~M.}\ \bibnamefont {Marcus}},\ }\bibfield  {title} {\enquote {\bibinfo
  {title} {Majorana bound state in a coupled quantum-dot hybrid-nanowire
  system},}\ }\href {\doibase 10.1126/science.aaf3961} {\bibfield  {journal}
  {\bibinfo  {journal} {Science}\ }\textbf {\bibinfo {volume} {354}},\ \bibinfo
  {pages} {1557--1562} (\bibinfo {year} {2016})}\BibitemShut {NoStop}%
\bibitem [{\citenamefont {Nadj-Perge}\ \emph {et~al.}(2014)\citenamefont
  {Nadj-Perge}, \citenamefont {Drozdov}, \citenamefont {Li}, \citenamefont
  {Chen}, \citenamefont {Jeon}, \citenamefont {Seo}, \citenamefont {MacDonald},
  \citenamefont {Bernevig},\ and\ \citenamefont {Yazdani}}]{NadjPerge2014}%
  \BibitemOpen
  \bibfield  {author} {\bibinfo {author} {\bibfnamefont {S.}~\bibnamefont
  {Nadj-Perge}}, \bibinfo {author} {\bibfnamefont {I.~K.}\ \bibnamefont
  {Drozdov}}, \bibinfo {author} {\bibfnamefont {J.}~\bibnamefont {Li}},
  \bibinfo {author} {\bibfnamefont {H.}~\bibnamefont {Chen}}, \bibinfo {author}
  {\bibfnamefont {S.}~\bibnamefont {Jeon}}, \bibinfo {author} {\bibfnamefont
  {J.}~\bibnamefont {Seo}}, \bibinfo {author} {\bibfnamefont {A.~H.}\
  \bibnamefont {MacDonald}}, \bibinfo {author} {\bibfnamefont {B.~A.}\
  \bibnamefont {Bernevig}}, \ and\ \bibinfo {author} {\bibfnamefont
  {A.}~\bibnamefont {Yazdani}},\ }\bibfield  {title} {\enquote {\bibinfo
  {title} {Observation of majorana fermions in ferromagnetic atomic chains on a
  superconductor},}\ }\href {\doibase 10.1126/science.1259327} {\bibfield
  {journal} {\bibinfo  {journal} {Science}\ }\textbf {\bibinfo {volume}
  {346}},\ \bibinfo {pages} {602--607} (\bibinfo {year} {2014})}\BibitemShut
  {NoStop}%
\bibitem [{\citenamefont {Ruby}\ \emph {et~al.}(2015)\citenamefont {Ruby},
  \citenamefont {Pientka}, \citenamefont {von Oppen}, \citenamefont
  {Heinrich},\ and\ \citenamefont {Franke}}]{Ruby2015}%
  \BibitemOpen
  \bibfield  {author} {\bibinfo {author} {\bibfnamefont {M.}~\bibnamefont
  {Ruby}}, \bibinfo {author} {\bibfnamefont {Y.}~\bibnamefont {Pientka},
  \bibfnamefont {F.and~Peng}}, \bibinfo {author} {\bibfnamefont
  {F.}~\bibnamefont {von Oppen}}, \bibinfo {author} {\bibfnamefont {B.~W.}\
  \bibnamefont {Heinrich}}, \ and\ \bibinfo {author} {\bibfnamefont {K.~J.}\
  \bibnamefont {Franke}},\ }\bibfield  {title} {\enquote {\bibinfo {title} {End
  states and subgap structure in proximity-coupled chains of magnetic
  adatoms},}\ }\href {\doibase 10.1103/PhysRevLett.115.197204} {\bibfield
  {journal} {\bibinfo  {journal} {Phys. Rev. Lett.}\ }\textbf {\bibinfo
  {volume} {115}},\ \bibinfo {pages} {197204} (\bibinfo {year}
  {2015})}\BibitemShut {NoStop}%
\bibitem [{\citenamefont {Pawlak}\ \emph {et~al.}(2016)\citenamefont {Pawlak},
  \citenamefont {Kisiel}, \citenamefont {Klinovaja}, \citenamefont {Meier},
  \citenamefont {Kawai}, \citenamefont {Glatzel}, \citenamefont {Loss},\ and\
  \citenamefont {Meyer}}]{Pawlak2016}%
  \BibitemOpen
  \bibfield  {author} {\bibinfo {author} {\bibfnamefont {R.}~\bibnamefont
  {Pawlak}}, \bibinfo {author} {\bibfnamefont {M.}~\bibnamefont {Kisiel}},
  \bibinfo {author} {\bibfnamefont {J.}~\bibnamefont {Klinovaja}}, \bibinfo
  {author} {\bibfnamefont {T.}~\bibnamefont {Meier}}, \bibinfo {author}
  {\bibfnamefont {S.}~\bibnamefont {Kawai}}, \bibinfo {author} {\bibfnamefont
  {T.}~\bibnamefont {Glatzel}}, \bibinfo {author} {\bibfnamefont
  {D.}~\bibnamefont {Loss}}, \ and\ \bibinfo {author} {\bibfnamefont
  {E.}~\bibnamefont {Meyer}},\ }\bibfield  {title} {\enquote {\bibinfo {title}
  {Probing atomic structure and majorana wavefunctions in mono-atomic fe chains
  on superconducting pb surface},}\ }\href
  {https://doi.org/10.1038/npjqi.2016.35} {\bibfield  {journal} {\bibinfo
  {journal} {Npj Quantum Information}\ }\textbf {\bibinfo {volume} {2}},\
  \bibinfo {pages} {16035 EP --} (\bibinfo {year} {2016})}\BibitemShut
  {NoStop}%
\bibitem [{\citenamefont {Zhang}\ \emph {et~al.}(2018)\citenamefont {Zhang},
  \citenamefont {Liu}, \citenamefont {Gazibegovic}, \citenamefont {Xu},
  \citenamefont {Logan}, \citenamefont {Wang}, \citenamefont {van Loo},
  \citenamefont {Bommer}, \citenamefont {de~Moor}, \citenamefont {Car},
  \citenamefont {Op~het Veld}, \citenamefont {van Veldhoven}, \citenamefont
  {Koelling}, \citenamefont {Verheijen}, \citenamefont {Pendharkar},
  \citenamefont {Pennachio}, \citenamefont {Shojaei}, \citenamefont {Lee},
  \citenamefont {Palmstr{\o}m}, \citenamefont {Bakkers}, \citenamefont
  {Sarma},\ and\ \citenamefont {Kouwenhoven}}]{Zhang2018}%
  \BibitemOpen
  \bibfield  {author} {\bibinfo {author} {\bibfnamefont {H.}~\bibnamefont
  {Zhang}}, \bibinfo {author} {\bibfnamefont {C.~X.}\ \bibnamefont {Liu}},
  \bibinfo {author} {\bibfnamefont {S.}~\bibnamefont {Gazibegovic}}, \bibinfo
  {author} {\bibfnamefont {D.}~\bibnamefont {Xu}}, \bibinfo {author}
  {\bibfnamefont {J.~A.}\ \bibnamefont {Logan}}, \bibinfo {author}
  {\bibfnamefont {G.}~\bibnamefont {Wang}}, \bibinfo {author} {\bibfnamefont
  {N.}~\bibnamefont {van Loo}}, \bibinfo {author} {\bibfnamefont {J.~D.~S.}\
  \bibnamefont {Bommer}}, \bibinfo {author} {\bibfnamefont {M.~W.~A.}\
  \bibnamefont {de~Moor}}, \bibinfo {author} {\bibfnamefont {D.}~\bibnamefont
  {Car}}, \bibinfo {author} {\bibfnamefont {R.L.~M.}\ \bibnamefont {Op~het
  Veld}}, \bibinfo {author} {\bibfnamefont {P.~J.}\ \bibnamefont {van
  Veldhoven}}, \bibinfo {author} {\bibfnamefont {S.}~\bibnamefont {Koelling}},
  \bibinfo {author} {\bibfnamefont {M.~A.}\ \bibnamefont {Verheijen}}, \bibinfo
  {author} {\bibfnamefont {M.}~\bibnamefont {Pendharkar}}, \bibinfo {author}
  {\bibfnamefont {D.~J.}\ \bibnamefont {Pennachio}}, \bibinfo {author}
  {\bibfnamefont {B.}~\bibnamefont {Shojaei}}, \bibinfo {author} {\bibfnamefont
  {J.~S.}\ \bibnamefont {Lee}}, \bibinfo {author} {\bibfnamefont {C.~J.}\
  \bibnamefont {Palmstr{\o}m}}, \bibinfo {author} {\bibfnamefont {E.~P. A.~M.}\
  \bibnamefont {Bakkers}}, \bibinfo {author} {\bibfnamefont {S.~Das}\
  \bibnamefont {Sarma}}, \ and\ \bibinfo {author} {\bibfnamefont {L.~P.}\
  \bibnamefont {Kouwenhoven}},\ }\bibfield  {title} {\enquote {\bibinfo {title}
  {Quantized majorana conductance},}\ }\href
  {https://doi.org/10.1038/nature26142} {\bibfield  {journal} {\bibinfo
  {journal} {Nature}\ }\textbf {\bibinfo {volume} {556}},\ \bibinfo {pages} {74
  EP --} (\bibinfo {year} {2018})}\BibitemShut {NoStop}%
\bibitem [{\citenamefont {Fornieri}\ \emph {et~al.}(2019)\citenamefont
  {Fornieri}, \citenamefont {Whiticar}, \citenamefont {Setiawan}, \citenamefont
  {Portol{\'e}s}, \citenamefont {Drachmann}, \citenamefont {Keselman},
  \citenamefont {Gronin}, \citenamefont {Thomas}, \citenamefont {Wang},
  \citenamefont {Kallaher}, \citenamefont {Gardner}, \citenamefont {Berg},
  \citenamefont {Manfra}, \citenamefont {Stern}, \citenamefont {Marcus},\ and\
  \citenamefont {Nichele}}]{Marcus2019}%
  \BibitemOpen
  \bibfield  {author} {\bibinfo {author} {\bibfnamefont {A.}~\bibnamefont
  {Fornieri}}, \bibinfo {author} {\bibfnamefont {A.~M.}\ \bibnamefont
  {Whiticar}}, \bibinfo {author} {\bibfnamefont {F.}~\bibnamefont {Setiawan}},
  \bibinfo {author} {\bibfnamefont {E.}~\bibnamefont {Portol{\'e}s}}, \bibinfo
  {author} {\bibfnamefont {A.~C.~C.}\ \bibnamefont {Drachmann}}, \bibinfo
  {author} {\bibfnamefont {A.}~\bibnamefont {Keselman}}, \bibinfo {author}
  {\bibfnamefont {S.}~\bibnamefont {Gronin}}, \bibinfo {author} {\bibfnamefont
  {C.}~\bibnamefont {Thomas}}, \bibinfo {author} {\bibfnamefont
  {T.}~\bibnamefont {Wang}}, \bibinfo {author} {\bibfnamefont {R.}~\bibnamefont
  {Kallaher}}, \bibinfo {author} {\bibfnamefont {G.~C.}\ \bibnamefont
  {Gardner}}, \bibinfo {author} {\bibfnamefont {E.}~\bibnamefont {Berg}},
  \bibinfo {author} {\bibfnamefont {M.~J.}\ \bibnamefont {Manfra}}, \bibinfo
  {author} {\bibfnamefont {A.}~\bibnamefont {Stern}}, \bibinfo {author}
  {\bibfnamefont {C.~M.}\ \bibnamefont {Marcus}}, \ and\ \bibinfo {author}
  {\bibfnamefont {F.}~\bibnamefont {Nichele}},\ }\bibfield  {title} {\enquote
  {\bibinfo {title} {Evidence of topological superconductivity in planar
  josephson junctions},}\ }\href {\doibase 10.1038/s41586-019-1068-8}
  {\bibfield  {journal} {\bibinfo  {journal} {Nature}\ }\textbf {\bibinfo
  {volume} {569}},\ \bibinfo {pages} {89--92} (\bibinfo {year}
  {2019})}\BibitemShut {NoStop}%
\bibitem [{\citenamefont {Ren}\ \emph {et~al.}(2019)\citenamefont {Ren},
  \citenamefont {Pientka}, \citenamefont {Hart}, \citenamefont {Pierce},
  \citenamefont {Kosowsky}, \citenamefont {Lunczer}, \citenamefont {Schlereth},
  \citenamefont {Scharf}, \citenamefont {Hankiewicz}, \citenamefont
  {Molenkamp}, \citenamefont {Halperin},\ and\ \citenamefont
  {Yacoby}}]{Falko2019}%
  \BibitemOpen
  \bibfield  {author} {\bibinfo {author} {\bibfnamefont {H.}~\bibnamefont
  {Ren}}, \bibinfo {author} {\bibfnamefont {F.}~\bibnamefont {Pientka}},
  \bibinfo {author} {\bibfnamefont {S.}~\bibnamefont {Hart}}, \bibinfo {author}
  {\bibfnamefont {A.~T.}\ \bibnamefont {Pierce}}, \bibinfo {author}
  {\bibfnamefont {M.}~\bibnamefont {Kosowsky}}, \bibinfo {author}
  {\bibfnamefont {L.}~\bibnamefont {Lunczer}}, \bibinfo {author} {\bibfnamefont
  {R.}~\bibnamefont {Schlereth}}, \bibinfo {author} {\bibfnamefont
  {B.}~\bibnamefont {Scharf}}, \bibinfo {author} {\bibfnamefont {E.~M.}\
  \bibnamefont {Hankiewicz}}, \bibinfo {author} {\bibfnamefont {L.~W.}\
  \bibnamefont {Molenkamp}}, \bibinfo {author} {\bibfnamefont {B.~I.}\
  \bibnamefont {Halperin}}, \ and\ \bibinfo {author} {\bibfnamefont
  {A.}~\bibnamefont {Yacoby}},\ }\bibfield  {title} {\enquote {\bibinfo {title}
  {Topological superconductivity in a phase-controlled josephson junction},}\
  }\href {\doibase 10.1038/s41586-019-1148-9} {\bibfield  {journal} {\bibinfo
  {journal} {Nature}\ }\textbf {\bibinfo {volume} {569}},\ \bibinfo {pages}
  {93--98} (\bibinfo {year} {2019})}\BibitemShut {NoStop}%
\bibitem [{\citenamefont {Rainis}\ and\ \citenamefont
  {Loss}(2012)}]{Rainis2012}%
  \BibitemOpen
  \bibfield  {author} {\bibinfo {author} {\bibfnamefont {D.}~\bibnamefont
  {Rainis}}\ and\ \bibinfo {author} {\bibfnamefont {D.}~\bibnamefont {Loss}},\
  }\bibfield  {title} {\enquote {\bibinfo {title} {Majorana qubit decoherence
  by quasiparticle poisoning},}\ }\href {\doibase 10.1103/PhysRevB.85.174533}
  {\bibfield  {journal} {\bibinfo  {journal} {Phys. Rev. B}\ }\textbf {\bibinfo
  {volume} {85}},\ \bibinfo {pages} {174533} (\bibinfo {year}
  {2012})}\BibitemShut {NoStop}%
\bibitem [{\citenamefont {Budich}\ \emph {et~al.}(2012)\citenamefont {Budich},
  \citenamefont {Walter},\ and\ \citenamefont {Trauzettel}}]{Budich2012}%
  \BibitemOpen
  \bibfield  {author} {\bibinfo {author} {\bibfnamefont {J.~C.}\ \bibnamefont
  {Budich}}, \bibinfo {author} {\bibfnamefont {S.}~\bibnamefont {Walter}}, \
  and\ \bibinfo {author} {\bibfnamefont {B.}~\bibnamefont {Trauzettel}},\
  }\bibfield  {title} {\enquote {\bibinfo {title} {Failure of protection of
  majorana based qubits against decoherence},}\ }\href {\doibase
  10.1103/PhysRevB.85.121405} {\bibfield  {journal} {\bibinfo  {journal} {Phys.
  Rev. B}\ }\textbf {\bibinfo {volume} {85}},\ \bibinfo {pages} {121405}
  (\bibinfo {year} {2012})}\BibitemShut {NoStop}%
\bibitem [{\citenamefont {Roy}\ \emph {et~al.}(2012)\citenamefont {Roy},
  \citenamefont {Bolech},\ and\ \citenamefont {Shah}}]{Roy2012}%
  \BibitemOpen
  \bibfield  {author} {\bibinfo {author} {\bibfnamefont {D.}~\bibnamefont
  {Roy}}, \bibinfo {author} {\bibfnamefont {C.~J.}\ \bibnamefont {Bolech}}, \
  and\ \bibinfo {author} {\bibfnamefont {N.}~\bibnamefont {Shah}},\ }\bibfield
  {title} {\enquote {\bibinfo {title} {Majorana fermions in a topological
  superconducting wire out of equilibrium: Exact microscopic transport analysis
  of a $p$-wave open chain coupled to normal leads},}\ }\href {\doibase
  10.1103/PhysRevB.86.094503} {\bibfield  {journal} {\bibinfo  {journal} {Phys.
  Rev. B}\ }\textbf {\bibinfo {volume} {86}},\ \bibinfo {pages} {094503}
  (\bibinfo {year} {2012})}\BibitemShut {NoStop}%
\bibitem [{\citenamefont {Konschelle}\ and\ \citenamefont
  {Hassler}(2013)}]{Konschelle2013}%
  \BibitemOpen
  \bibfield  {author} {\bibinfo {author} {\bibfnamefont {F.}~\bibnamefont
  {Konschelle}}\ and\ \bibinfo {author} {\bibfnamefont {F.}~\bibnamefont
  {Hassler}},\ }\bibfield  {title} {\enquote {\bibinfo {title} {Effects of
  nonequilibrium noise on a quantum memory encoded in majorana zero modes},}\
  }\href {\doibase 10.1103/PhysRevB.88.075431} {\bibfield  {journal} {\bibinfo
  {journal} {Phys. Rev. B}\ }\textbf {\bibinfo {volume} {88}},\ \bibinfo
  {pages} {075431} (\bibinfo {year} {2013})}\BibitemShut {NoStop}%
\bibitem [{\citenamefont {Yang}\ and\ \citenamefont
  {Feldman}(2014)}]{Yang2014}%
  \BibitemOpen
  \bibfield  {author} {\bibinfo {author} {\bibfnamefont {G.}~\bibnamefont
  {Yang}}\ and\ \bibinfo {author} {\bibfnamefont {D.~E.}\ \bibnamefont
  {Feldman}},\ }\bibfield  {title} {\enquote {\bibinfo {title} {Exact zero
  modes and decoherence in systems of interacting majorana fermions},}\ }\href
  {\doibase 10.1103/PhysRevB.89.035136} {\bibfield  {journal} {\bibinfo
  {journal} {Phys. Rev. B}\ }\textbf {\bibinfo {volume} {89}},\ \bibinfo
  {pages} {035136} (\bibinfo {year} {2014})}\BibitemShut {NoStop}%
\bibitem [{\citenamefont {Ng}(2015)}]{Ng2015}%
  \BibitemOpen
  \bibfield  {author} {\bibinfo {author} {\bibfnamefont {H.~T.}\ \bibnamefont
  {Ng}},\ }\bibfield  {title} {\enquote {\bibinfo {title} {Decoherence of
  interacting majorana modes},}\ }\href@noop {} {\bibfield  {journal} {\bibinfo
   {journal} {Scientific Reports}\ }\textbf {\bibinfo {volume} {5}},\ \bibinfo
  {pages} {12530 EP --} (\bibinfo {year} {2015})}\BibitemShut {NoStop}%
\bibitem [{\citenamefont {Pedrocchi}\ and\ \citenamefont
  {DiVincenzo}(2015)}]{Pedrocchi2015}%
  \BibitemOpen
  \bibfield  {author} {\bibinfo {author} {\bibfnamefont {F.~L.}\ \bibnamefont
  {Pedrocchi}}\ and\ \bibinfo {author} {\bibfnamefont {D.~P.}\ \bibnamefont
  {DiVincenzo}},\ }\bibfield  {title} {\enquote {\bibinfo {title} {Majorana
  braiding with thermal noise},}\ }\href {\doibase
  10.1103/PhysRevLett.115.120402} {\bibfield  {journal} {\bibinfo  {journal}
  {Phys. Rev. Lett.}\ }\textbf {\bibinfo {volume} {115}},\ \bibinfo {pages}
  {120402} (\bibinfo {year} {2015})}\BibitemShut {NoStop}%
\bibitem [{\citenamefont {Hu}\ \emph {et~al.}(2015)\citenamefont {Hu},
  \citenamefont {Cai}, \citenamefont {Baranov},\ and\ \citenamefont
  {Zoller}}]{Hu2015}%
  \BibitemOpen
  \bibfield  {author} {\bibinfo {author} {\bibfnamefont {Y.}~\bibnamefont
  {Hu}}, \bibinfo {author} {\bibfnamefont {Z.}~\bibnamefont {Cai}}, \bibinfo
  {author} {\bibfnamefont {M.~A.}\ \bibnamefont {Baranov}}, \ and\ \bibinfo
  {author} {\bibfnamefont {P.}~\bibnamefont {Zoller}},\ }\bibfield  {title}
  {\enquote {\bibinfo {title} {Majorana fermions in noisy kitaev wires},}\
  }\href {\doibase 10.1103/PhysRevB.92.165118} {\bibfield  {journal} {\bibinfo
  {journal} {Phys. Rev. B}\ }\textbf {\bibinfo {volume} {92}},\ \bibinfo
  {pages} {165118} (\bibinfo {year} {2015})}\BibitemShut {NoStop}%
\bibitem [{\citenamefont {Ippoliti}\ \emph {et~al.}(2016)\citenamefont
  {Ippoliti}, \citenamefont {Rizzi}, \citenamefont {Giovannetti},\ and\
  \citenamefont {Mazza}}]{Ippoliti2016}%
  \BibitemOpen
  \bibfield  {author} {\bibinfo {author} {\bibfnamefont {M.}~\bibnamefont
  {Ippoliti}}, \bibinfo {author} {\bibfnamefont {M.}~\bibnamefont {Rizzi}},
  \bibinfo {author} {\bibfnamefont {V.}~\bibnamefont {Giovannetti}}, \ and\
  \bibinfo {author} {\bibfnamefont {L.}~\bibnamefont {Mazza}},\ }\bibfield
  {title} {\enquote {\bibinfo {title} {Quantum memories with zero-energy
  majorana modes and experimental constraints},}\ }\href {\doibase
  10.1103/PhysRevA.93.062325} {\bibfield  {journal} {\bibinfo  {journal} {Phys.
  Rev. A}\ }\textbf {\bibinfo {volume} {93}},\ \bibinfo {pages} {062325}
  (\bibinfo {year} {2016})}\BibitemShut {NoStop}%
\bibitem [{\citenamefont {Brown}\ \emph {et~al.}(2016)\citenamefont {Brown},
  \citenamefont {Loss}, \citenamefont {Pachos}, \citenamefont {Self},\ and\
  \citenamefont {Wootton}}]{Brown2016}%
  \BibitemOpen
  \bibfield  {author} {\bibinfo {author} {\bibfnamefont {B.~J.}\ \bibnamefont
  {Brown}}, \bibinfo {author} {\bibfnamefont {D.}~\bibnamefont {Loss}},
  \bibinfo {author} {\bibfnamefont {J.~K.}\ \bibnamefont {Pachos}}, \bibinfo
  {author} {\bibfnamefont {C.~N.}\ \bibnamefont {Self}}, \ and\ \bibinfo
  {author} {\bibfnamefont {J.~R.}\ \bibnamefont {Wootton}},\ }\bibfield
  {title} {\enquote {\bibinfo {title} {Quantum memories at finite
  temperature},}\ }\href {\doibase 10.1103/RevModPhys.88.045005} {\bibfield
  {journal} {\bibinfo  {journal} {Rev. Mod. Phys.}\ }\textbf {\bibinfo {volume}
  {88}},\ \bibinfo {pages} {045005} (\bibinfo {year} {2016})}\BibitemShut
  {NoStop}%
\bibitem [{\citenamefont {Aseev}\ \emph {et~al.}(2018)\citenamefont {Aseev},
  \citenamefont {Klinovaja},\ and\ \citenamefont {Loss}}]{Aseev2018}%
  \BibitemOpen
  \bibfield  {author} {\bibinfo {author} {\bibfnamefont {P.~P.}\ \bibnamefont
  {Aseev}}, \bibinfo {author} {\bibfnamefont {J.}~\bibnamefont {Klinovaja}}, \
  and\ \bibinfo {author} {\bibfnamefont {D.}~\bibnamefont {Loss}},\ }\bibfield
  {title} {\enquote {\bibinfo {title} {Lifetime of majorana qubits in rashba
  nanowires with nonuniform chemical potential},}\ }\href {\doibase
  10.1103/PhysRevB.98.155414} {\bibfield  {journal} {\bibinfo  {journal} {Phys.
  Rev. B}\ }\textbf {\bibinfo {volume} {98}},\ \bibinfo {pages} {155414}
  (\bibinfo {year} {2018})}\BibitemShut {NoStop}%
\bibitem [{\citenamefont {Knapp}\ \emph {et~al.}(2018)\citenamefont {Knapp},
  \citenamefont {Karzig}, \citenamefont {Lutchyn},\ and\ \citenamefont
  {Nayak}}]{Knapp2018}%
  \BibitemOpen
  \bibfield  {author} {\bibinfo {author} {\bibfnamefont {C.}~\bibnamefont
  {Knapp}}, \bibinfo {author} {\bibfnamefont {T.}~\bibnamefont {Karzig}},
  \bibinfo {author} {\bibfnamefont {R.~M.}\ \bibnamefont {Lutchyn}}, \ and\
  \bibinfo {author} {\bibfnamefont {C.}~\bibnamefont {Nayak}},\ }\bibfield
  {title} {\enquote {\bibinfo {title} {Dephasing of majorana-based qubits},}\
  }\href {\doibase 10.1103/PhysRevB.97.125404} {\bibfield  {journal} {\bibinfo
  {journal} {Phys. Rev. B}\ }\textbf {\bibinfo {volume} {97}},\ \bibinfo
  {pages} {125404} (\bibinfo {year} {2018})}\BibitemShut {NoStop}%
\bibitem [{\citenamefont {Zhang}\ \emph {et~al.}(2019)\citenamefont {Zhang},
  \citenamefont {Mei}, \citenamefont {Meng}, \citenamefont {Liang},\ and\
  \citenamefont {Yang}}]{Zhang2019}%
  \BibitemOpen
  \bibfield  {author} {\bibinfo {author} {\bibfnamefont {Z.~T.}\ \bibnamefont
  {Zhang}}, \bibinfo {author} {\bibfnamefont {F.}~\bibnamefont {Mei}}, \bibinfo
  {author} {\bibfnamefont {X.~G.}\ \bibnamefont {Meng}}, \bibinfo {author}
  {\bibfnamefont {B.~L.}\ \bibnamefont {Liang}}, \ and\ \bibinfo {author}
  {\bibfnamefont {Z.~S.}\ \bibnamefont {Yang}},\ }\bibfield  {title} {\enquote
  {\bibinfo {title} {Effects of decoherence on diabatic errors in majorana
  braiding},}\ }\href {https://arxiv.org/abs/1902.05807} {\bibfield  {journal}
  {\bibinfo  {journal} {Journal of Mathematical Physics}\ } (\bibinfo {year}
  {2019})},\ \Eprint {http://arxiv.org/abs/arXiv:1902.05807} {arXiv:1902.05807}
  \BibitemShut {NoStop}%
\bibitem [{\citenamefont {Karzig}\ \emph {et~al.}(2013)\citenamefont {Karzig},
  \citenamefont {Refael},\ and\ \citenamefont {von Oppen}}]{Karzig2013}%
  \BibitemOpen
  \bibfield  {author} {\bibinfo {author} {\bibfnamefont {T.}~\bibnamefont
  {Karzig}}, \bibinfo {author} {\bibfnamefont {G.}~\bibnamefont {Refael}}, \
  and\ \bibinfo {author} {\bibfnamefont {F.}~\bibnamefont {von Oppen}},\
  }\bibfield  {title} {\enquote {\bibinfo {title} {Boosting majorana zero
  modes},}\ }\href {\doibase 10.1103/PhysRevX.3.041017} {\bibfield  {journal}
  {\bibinfo  {journal} {Phys. Rev. X}\ }\textbf {\bibinfo {volume} {3}},\
  \bibinfo {pages} {041017} (\bibinfo {year} {2013})}\BibitemShut {NoStop}%
\bibitem [{\citenamefont {Scheurer}\ and\ \citenamefont
  {Shnirman}(2013)}]{Scheurer2013}%
  \BibitemOpen
  \bibfield  {author} {\bibinfo {author} {\bibfnamefont {M.~S.}\ \bibnamefont
  {Scheurer}}\ and\ \bibinfo {author} {\bibfnamefont {A.}~\bibnamefont
  {Shnirman}},\ }\bibfield  {title} {\enquote {\bibinfo {title} {Nonadiabatic
  processes in majorana qubit systems},}\ }\href {\doibase
  10.1103/PhysRevB.88.064515} {\bibfield  {journal} {\bibinfo  {journal} {Phys.
  Rev. B}\ }\textbf {\bibinfo {volume} {88}},\ \bibinfo {pages} {064515}
  (\bibinfo {year} {2013})}\BibitemShut {NoStop}%
\bibitem [{\citenamefont {Coopmans}\ \emph {et~al.}(2021)\citenamefont
  {Coopmans}, \citenamefont {Luo}, \citenamefont {Kells}, \citenamefont
  {Clark},\ and\ \citenamefont {Carrasquilla}}]{coopmans2020protocol}%
  \BibitemOpen
  \bibfield  {author} {\bibinfo {author} {\bibfnamefont {L.}~\bibnamefont
  {Coopmans}}, \bibinfo {author} {\bibfnamefont {D.}~\bibnamefont {Luo}},
  \bibinfo {author} {\bibfnamefont {G.}~\bibnamefont {Kells}}, \bibinfo
  {author} {\bibfnamefont {B.~K.}\ \bibnamefont {Clark}}, \ and\ \bibinfo
  {author} {\bibfnamefont {J.}~\bibnamefont {Carrasquilla}},\ }\bibfield
  {title} {\enquote {\bibinfo {title} {Protocol discovery for the quantum
  control of majoranas by differentiable programming and natural evolution
  strategies},}\ }\href {\doibase 10.1103/PRXQuantum.2.020332} {\bibfield
  {journal} {\bibinfo  {journal} {PRX Quantum}\ }\textbf {\bibinfo {volume}
  {2}},\ \bibinfo {pages} {020332} (\bibinfo {year} {2021})}\BibitemShut
  {NoStop}%
\bibitem [{\citenamefont {Moon}\ and\ \citenamefont
  {Nachtergaele}(2018)}]{Moon2018}%
  \BibitemOpen
  \bibfield  {author} {\bibinfo {author} {\bibfnamefont {A.}~\bibnamefont
  {Moon}}\ and\ \bibinfo {author} {\bibfnamefont {B.}~\bibnamefont
  {Nachtergaele}},\ }\bibfield  {title} {\enquote {\bibinfo {title} {Stability
  of gapped ground state phases of spins and fermions in one dimension},}\
  }\href {\doibase 10.1063/1.5036751} {\bibfield  {journal} {\bibinfo
  {journal} {Journal of Mathematical Physics}\ }\textbf {\bibinfo {volume}
  {59}},\ \bibinfo {pages} {091415} (\bibinfo {year} {2018})},\ \Eprint
  {http://arxiv.org/abs/https://doi.org/10.1063/1.5036751}
  {https://doi.org/10.1063/1.5036751} \BibitemShut {NoStop}%
\bibitem [{\citenamefont {Kells}(2015{\natexlab{a}})}]{Kells2015}%
  \BibitemOpen
  \bibfield  {author} {\bibinfo {author} {\bibfnamefont {G.}~\bibnamefont
  {Kells}},\ }\bibfield  {title} {\enquote {\bibinfo {title} {Many-body
  majorana operators and the equivalence of parity sectors},}\ }\href {\doibase
  10.1103/PhysRevB.92.081401} {\bibfield  {journal} {\bibinfo  {journal} {Phys.
  Rev. B}\ }\textbf {\bibinfo {volume} {92}},\ \bibinfo {pages} {081401}
  (\bibinfo {year} {2015}{\natexlab{a}})}\BibitemShut {NoStop}%
\bibitem [{\citenamefont {Moran}\ \emph {et~al.}(2017)\citenamefont {Moran},
  \citenamefont {Pellegrino}, \citenamefont {Slingerland},\ and\ \citenamefont
  {Kells}}]{Moran2017}%
  \BibitemOpen
  \bibfield  {author} {\bibinfo {author} {\bibfnamefont {N.}~\bibnamefont
  {Moran}}, \bibinfo {author} {\bibfnamefont {D.}~\bibnamefont {Pellegrino}},
  \bibinfo {author} {\bibfnamefont {J.~K.}\ \bibnamefont {Slingerland}}, \ and\
  \bibinfo {author} {\bibfnamefont {G.}~\bibnamefont {Kells}},\ }\bibfield
  {title} {\enquote {\bibinfo {title} {Parafermionic clock models and quantum
  resonance},}\ }\href {\doibase 10.1103/PhysRevB.95.235127} {\bibfield
  {journal} {\bibinfo  {journal} {Phys. Rev. B}\ }\textbf {\bibinfo {volume}
  {95}},\ \bibinfo {pages} {235127} (\bibinfo {year} {2017})}\BibitemShut
  {NoStop}%
\bibitem [{\citenamefont {Kells}\ \emph {et~al.}(2018)\citenamefont {Kells},
  \citenamefont {Moran},\ and\ \citenamefont {Meidan}}]{Kells2018b}%
  \BibitemOpen
  \bibfield  {author} {\bibinfo {author} {\bibfnamefont {G.}~\bibnamefont
  {Kells}}, \bibinfo {author} {\bibfnamefont {N.}~\bibnamefont {Moran}}, \ and\
  \bibinfo {author} {\bibfnamefont {D.}~\bibnamefont {Meidan}},\ }\bibfield
  {title} {\enquote {\bibinfo {title} {Localization enhanced and degraded
  topological order in interacting $p$-wave wires},}\ }\href {\doibase
  10.1103/PhysRevB.97.085425} {\bibfield  {journal} {\bibinfo  {journal} {Phys.
  Rev. B}\ }\textbf {\bibinfo {volume} {97}},\ \bibinfo {pages} {085425}
  (\bibinfo {year} {2018})}\BibitemShut {NoStop}%
\bibitem [{\citenamefont {Pellegrino}\ \emph {et~al.}(2020)\citenamefont
  {Pellegrino}, \citenamefont {Kells}, \citenamefont {Moran},\ and\
  \citenamefont {Slingerland}}]{Pellegrino2020}%
  \BibitemOpen
  \bibfield  {author} {\bibinfo {author} {\bibfnamefont {D}~\bibnamefont
  {Pellegrino}}, \bibinfo {author} {\bibfnamefont {G}~\bibnamefont {Kells}},
  \bibinfo {author} {\bibfnamefont {N}~\bibnamefont {Moran}}, \ and\ \bibinfo
  {author} {\bibfnamefont {J~K}\ \bibnamefont {Slingerland}},\ }\bibfield
  {title} {\enquote {\bibinfo {title} {Constructing edge zero modes through
  domain wall angle conservation},}\ }\href {\doibase 10.1088/1751-8121/ab6fc7}
  {\bibfield  {journal} {\bibinfo  {journal} {Journal of Physics A:
  Mathematical and Theoretical}\ }\textbf {\bibinfo {volume} {53}},\ \bibinfo
  {pages} {095006} (\bibinfo {year} {2020})}\BibitemShut {NoStop}%
\bibitem [{Note5()}]{Note5}%
  \BibitemOpen
  \bibinfo {note} {The ground state energy splitting is time-averaged because
  the spectrum to varying in time due to the time-dependent
  perturbation.}\BibitemShut {Stop}%
\bibitem [{Note6()}]{Note6}%
  \BibitemOpen
  \bibinfo {note} {For a box function, with cut-off $T^{*}$, $\protect \mathcal
  {F}[\Theta ]$ has the form of a sinc function of width $\sim 1/T^*$, and
  hence a convolution with $\protect \mathcal {F}[\Theta ]$ acts to smear
  $\protect \mathcal {F}[G^a](k,\omega )$ by this width in the $\omega $
  direction.}\BibitemShut {Stop}%
\bibitem [{\citenamefont {{Nachtergaele}}\ \emph {et~al.}(2018)\citenamefont
  {{Nachtergaele}}, \citenamefont {{Sims}},\ and\ \citenamefont
  {{Young}}}]{2017arXiv170508553N}%
  \BibitemOpen
  \bibfield  {author} {\bibinfo {author} {\bibfnamefont {B.}~\bibnamefont
  {{Nachtergaele}}}, \bibinfo {author} {\bibfnamefont {R.}~\bibnamefont
  {{Sims}}}, \ and\ \bibinfo {author} {\bibfnamefont {A.}~\bibnamefont
  {{Young}}},\ }\bibfield  {title} {\enquote {\bibinfo {title} {Lieb-robinson
  bounds, the spectral flow, and stability of the spectral gap for lattice
  fermion systems},}\ }\href {\doibase 10.1090/conm/717} {\bibfield  {journal}
  {\bibinfo  {journal} {Contemporary Mathematics}\ } (\bibinfo {year} {2018}),\
  10.1090/conm/717}\BibitemShut {NoStop}%
\bibitem [{\citenamefont {Gangadharaiah}\ \emph {et~al.}(2011)\citenamefont
  {Gangadharaiah}, \citenamefont {Braunecker}, \citenamefont {Simon},\ and\
  \citenamefont {Loss}}]{Gangadharaiah2011}%
  \BibitemOpen
  \bibfield  {author} {\bibinfo {author} {\bibfnamefont {S.}~\bibnamefont
  {Gangadharaiah}}, \bibinfo {author} {\bibfnamefont {B.}~\bibnamefont
  {Braunecker}}, \bibinfo {author} {\bibfnamefont {P.}~\bibnamefont {Simon}}, \
  and\ \bibinfo {author} {\bibfnamefont {D.}~\bibnamefont {Loss}},\ }\bibfield
  {title} {\enquote {\bibinfo {title} {Majorana edge states in interacting
  one-dimensional systems},}\ }\href {\doibase 10.1103/PhysRevLett.107.036801}
  {\bibfield  {journal} {\bibinfo  {journal} {Phys. Rev. Lett.}\ }\textbf
  {\bibinfo {volume} {107}},\ \bibinfo {pages} {036801} (\bibinfo {year}
  {2011})}\BibitemShut {NoStop}%
\bibitem [{\citenamefont {Jermyn}\ \emph {et~al.}(2014)\citenamefont {Jermyn},
  \citenamefont {Mong}, \citenamefont {Alicea},\ and\ \citenamefont
  {Fendley}}]{Jermyn2014}%
  \BibitemOpen
  \bibfield  {author} {\bibinfo {author} {\bibfnamefont {A.~S.}\ \bibnamefont
  {Jermyn}}, \bibinfo {author} {\bibfnamefont {R.~S.~K.}\ \bibnamefont {Mong}},
  \bibinfo {author} {\bibfnamefont {J.}~\bibnamefont {Alicea}}, \ and\ \bibinfo
  {author} {\bibfnamefont {P.}~\bibnamefont {Fendley}},\ }\bibfield  {title}
  {\enquote {\bibinfo {title} {Stability of zero modes in parafermion
  chains},}\ }\href {\doibase 10.1103/PhysRevB.90.165106} {\bibfield  {journal}
  {\bibinfo  {journal} {Phys. Rev. B}\ }\textbf {\bibinfo {volume} {90}},\
  \bibinfo {pages} {165106} (\bibinfo {year} {2014})}\BibitemShut {NoStop}%
\bibitem [{\citenamefont {Kells}(2015{\natexlab{b}})}]{Kells2015b}%
  \BibitemOpen
  \bibfield  {author} {\bibinfo {author} {\bibfnamefont {G.}~\bibnamefont
  {Kells}},\ }\bibfield  {title} {\enquote {\bibinfo {title} {Multiparticle
  content of majorana zero modes in the interacting $p$-wave wire},}\ }\href
  {\doibase 10.1103/PhysRevB.92.155434} {\bibfield  {journal} {\bibinfo
  {journal} {Phys. Rev. B}\ }\textbf {\bibinfo {volume} {92}},\ \bibinfo
  {pages} {155434} (\bibinfo {year} {2015}{\natexlab{b}})}\BibitemShut
  {NoStop}%
\bibitem [{\citenamefont {Fendley}(2016)}]{Fendley2016}%
  \BibitemOpen
  \bibfield  {author} {\bibinfo {author} {\bibfnamefont {P.}~\bibnamefont
  {Fendley}},\ }\bibfield  {title} {\enquote {\bibinfo {title} {Strong zero
  modes and eigenstate phase transitions in the {XYZ}/interacting majorana
  chain},}\ }\href {\doibase 10.1088/1751-8113/49/30/30lt01} {\bibfield
  {journal} {\bibinfo  {journal} {Journal of Physics A: Mathematical and
  Theoretical}\ }\textbf {\bibinfo {volume} {49}},\ \bibinfo {pages} {30LT01}
  (\bibinfo {year} {2016})}\BibitemShut {NoStop}%
\bibitem [{\citenamefont {Kemp}\ \emph {et~al.}(2017)\citenamefont {Kemp},
  \citenamefont {Yao}, \citenamefont {Laumann},\ and\ \citenamefont
  {Fendley}}]{Kemp2017}%
  \BibitemOpen
  \bibfield  {author} {\bibinfo {author} {\bibfnamefont {J.}~\bibnamefont
  {Kemp}}, \bibinfo {author} {\bibfnamefont {N.~Y.}\ \bibnamefont {Yao}},
  \bibinfo {author} {\bibfnamefont {C.~R.}\ \bibnamefont {Laumann}}, \ and\
  \bibinfo {author} {\bibfnamefont {P.}~\bibnamefont {Fendley}},\ }\bibfield
  {title} {\enquote {\bibinfo {title} {Long coherence times for edge spins},}\
  }\href {\doibase 10.1088/1742-5468/aa73f0} {\bibfield  {journal} {\bibinfo
  {journal} {Journal of Statistical Mechanics: Theory and Experiment}\ }\textbf
  {\bibinfo {volume} {2017}},\ \bibinfo {pages} {063105} (\bibinfo {year}
  {2017})}\BibitemShut {NoStop}%
\bibitem [{\citenamefont {McGinley}\ \emph {et~al.}(2017)\citenamefont
  {McGinley}, \citenamefont {Knolle},\ and\ \citenamefont
  {Nunnenkamp}}]{McGinley2017}%
  \BibitemOpen
  \bibfield  {author} {\bibinfo {author} {\bibfnamefont {M.}~\bibnamefont
  {McGinley}}, \bibinfo {author} {\bibfnamefont {J.}~\bibnamefont {Knolle}}, \
  and\ \bibinfo {author} {\bibfnamefont {A.}~\bibnamefont {Nunnenkamp}},\
  }\bibfield  {title} {\enquote {\bibinfo {title} {Robustness of majorana edge
  modes and topological order: Exact results for the symmetric interacting
  kitaev chain with disorder},}\ }\href {\doibase 10.1103/PhysRevB.96.241113}
  {\bibfield  {journal} {\bibinfo  {journal} {Phys. Rev. B}\ }\textbf {\bibinfo
  {volume} {96}},\ \bibinfo {pages} {241113} (\bibinfo {year}
  {2017})}\BibitemShut {NoStop}%
\bibitem [{\citenamefont {Mahyaeh}\ and\ \citenamefont
  {Ardonne}(2018)}]{Ardonne2018}%
  \BibitemOpen
  \bibfield  {author} {\bibinfo {author} {\bibfnamefont {I.}~\bibnamefont
  {Mahyaeh}}\ and\ \bibinfo {author} {\bibfnamefont {E.}~\bibnamefont
  {Ardonne}},\ }\bibfield  {title} {\enquote {\bibinfo {title} {Exact results
  for a ${\mathbb{z}}_{3}$-clock-type model and some close relatives},}\ }\href
  {\doibase 10.1103/PhysRevB.98.245104} {\bibfield  {journal} {\bibinfo
  {journal} {Phys. Rev. B}\ }\textbf {\bibinfo {volume} {98}},\ \bibinfo
  {pages} {245104} (\bibinfo {year} {2018})}\BibitemShut {NoStop}%
\bibitem [{\citenamefont {Mahyaeh}\ and\ \citenamefont
  {Ardonne}(2020)}]{Ardonne2020}%
  \BibitemOpen
  \bibfield  {author} {\bibinfo {author} {\bibfnamefont {I.}~\bibnamefont
  {Mahyaeh}}\ and\ \bibinfo {author} {\bibfnamefont {E.}~\bibnamefont
  {Ardonne}},\ }\bibfield  {title} {\enquote {\bibinfo {title} {Study of the
  phase diagram of the kitaev-hubbard chain},}\ }\href {\doibase
  10.1103/PhysRevB.101.085125} {\bibfield  {journal} {\bibinfo  {journal}
  {Phys. Rev. B}\ }\textbf {\bibinfo {volume} {101}},\ \bibinfo {pages}
  {085125} (\bibinfo {year} {2020})}\BibitemShut {NoStop}%
\bibitem [{\citenamefont {Bradlyn}\ \emph {et~al.}(2017)\citenamefont
  {Bradlyn}, \citenamefont {Elcoro}, \citenamefont {Cano}, \citenamefont
  {Vergniory}, \citenamefont {Wang}, \citenamefont {Felser}, \citenamefont
  {Aroyo},\ and\ \citenamefont {Bernevig}}]{Bradlyn2017}%
  \BibitemOpen
  \bibfield  {author} {\bibinfo {author} {\bibfnamefont {B.}~\bibnamefont
  {Bradlyn}}, \bibinfo {author} {\bibfnamefont {L.}~\bibnamefont {Elcoro}},
  \bibinfo {author} {\bibfnamefont {J.}~\bibnamefont {Cano}}, \bibinfo {author}
  {\bibfnamefont {M.~G.}\ \bibnamefont {Vergniory}}, \bibinfo {author}
  {\bibfnamefont {Z.}~\bibnamefont {Wang}}, \bibinfo {author} {\bibfnamefont
  {C.}~\bibnamefont {Felser}}, \bibinfo {author} {\bibfnamefont {M.~I.}\
  \bibnamefont {Aroyo}}, \ and\ \bibinfo {author} {\bibfnamefont {B.~A.}\
  \bibnamefont {Bernevig}},\ }\bibfield  {title} {\enquote {\bibinfo {title}
  {Topological quantum chemistry},}\ }\href {\doibase 10.1038/nature23268}
  {\bibfield  {journal} {\bibinfo  {journal} {Nature}\ }\textbf {\bibinfo
  {volume} {547}},\ \bibinfo {pages} {298--305} (\bibinfo {year}
  {2017})}\BibitemShut {NoStop}%
\bibitem [{\citenamefont {Huse}\ \emph {et~al.}(2013)\citenamefont {Huse},
  \citenamefont {Nandkishore}, \citenamefont {Oganesyan}, \citenamefont {Pal},\
  and\ \citenamefont {Sondhi}}]{Huse2013}%
  \BibitemOpen
  \bibfield  {author} {\bibinfo {author} {\bibfnamefont {D.~A.}\ \bibnamefont
  {Huse}}, \bibinfo {author} {\bibfnamefont {R.}~\bibnamefont {Nandkishore}},
  \bibinfo {author} {\bibfnamefont {V.}~\bibnamefont {Oganesyan}}, \bibinfo
  {author} {\bibfnamefont {A.}~\bibnamefont {Pal}}, \ and\ \bibinfo {author}
  {\bibfnamefont {S.~L.}\ \bibnamefont {Sondhi}},\ }\bibfield  {title}
  {\enquote {\bibinfo {title} {Localization-protected quantum order},}\ }\href
  {\doibase 10.1103/physrevb.88.014206} {\bibfield  {journal} {\bibinfo
  {journal} {Physical Review B}\ }\textbf {\bibinfo {volume} {88}} (\bibinfo
  {year} {2013}),\ 10.1103/physrevb.88.014206}\BibitemShut {NoStop}%
\bibitem [{\citenamefont {Else}\ \emph {et~al.}(2017)\citenamefont {Else},
  \citenamefont {Fendley}, \citenamefont {Kemp},\ and\ \citenamefont
  {Nayak}}]{Else2017}%
  \BibitemOpen
  \bibfield  {author} {\bibinfo {author} {\bibfnamefont {D.~V.}\ \bibnamefont
  {Else}}, \bibinfo {author} {\bibfnamefont {P.}~\bibnamefont {Fendley}},
  \bibinfo {author} {\bibfnamefont {J.}~\bibnamefont {Kemp}}, \ and\ \bibinfo
  {author} {\bibfnamefont {C.}~\bibnamefont {Nayak}},\ }\bibfield  {title}
  {\enquote {\bibinfo {title} {Prethermal strong zero modes and topological
  qubits},}\ }\href {\doibase 10.1103/PhysRevX.7.041062} {\bibfield  {journal}
  {\bibinfo  {journal} {Phys. Rev. X}\ }\textbf {\bibinfo {volume} {7}},\
  \bibinfo {pages} {041062} (\bibinfo {year} {2017})}\BibitemShut {NoStop}%
\bibitem [{\citenamefont {Nulty}\ \emph {et~al.}(2020)\citenamefont {Nulty},
  \citenamefont {Vala}, \citenamefont {Meidan},\ and\ \citenamefont
  {Kells}}]{Nulty2020}%
  \BibitemOpen
  \bibfield  {author} {\bibinfo {author} {\bibfnamefont {S.}~\bibnamefont
  {Nulty}}, \bibinfo {author} {\bibfnamefont {J.}~\bibnamefont {Vala}},
  \bibinfo {author} {\bibfnamefont {D.}~\bibnamefont {Meidan}}, \ and\ \bibinfo
  {author} {\bibfnamefont {G.}~\bibnamefont {Kells}},\ }\bibfield  {title}
  {\enquote {\bibinfo {title} {Constrained thermalization and topological
  superconductivity},}\ }\href {\doibase 10.1103/PhysRevB.102.054508}
  {\bibfield  {journal} {\bibinfo  {journal} {Phys. Rev. B}\ }\textbf {\bibinfo
  {volume} {102}},\ \bibinfo {pages} {054508} (\bibinfo {year}
  {2020})}\BibitemShut {NoStop}%
\end{thebibliography}%
\end{document}